\documentclass[conference]{IEEEtran}
\pdfoutput=1
\ifCLASSINFOpdf
   \usepackage[pdftex]{graphicx}
\else
\fi
\begin{document}
\title{A Case Study in Matching Service Descriptions to Implementations in an
Existing System}

\author{\IEEEauthorblockN{Hari S. Gupta\IEEEauthorrefmark{1},
Deepak D'Souza\IEEEauthorrefmark{1},
Raghavan Komondoor\IEEEauthorrefmark{1} and 
Girish M. Rama\IEEEauthorrefmark{2}}
\IEEEauthorblockA{\IEEEauthorrefmark{1}Indian Institute of Science,
Bangalore, India\\ Email: hari.csa.iisc@gmail.com,\{deepakd, raghavan\}@csa.iisc.ernet.in}
\IEEEauthorblockA{\IEEEauthorrefmark{2}Infosys Technologies Ltd., India\\
Email: Girish\_Rama@infosys.com}}

\maketitle

 \begin{abstract}
 A number of companies are trying to migrate large monolithic software
 systems to  Service Oriented Architectures.  A common approach to do this
 is to first identify and describe desired services (i.e., create a model),
 and then to locate portions of code within the existing system that
 implement the described services. In this paper we describe a detailed case
 study we undertook to match a model to an open-source business application.
 We describe the systematic methodology we used, the results of the
 exercise, as well as several observations that throw light on the nature of
 this problem. We  also suggest and validate heuristics that are likely to
 be useful in partially automating the process of matching service
 descriptions to implementations in existing applications.
 \end{abstract}

\IEEEpeerreviewmaketitle

\section{Introduction}

A large number of organizations are saddled with monolithic applications that
combine too many different independent functionalities. Many of these
organizations are in the process of migrating these applications to a
Service Oriented Architecture
(SOA)~\cite{kontogiannis2008research, papazoglou2008service}, with the goal
of improving the maintainability of the applications, the reuse-ability of
individual functionalities within the applications, and the potential for
integration with other applications.  A common approach to this is to start
with an analysis of the business domain, identify the important business
processes, and use these to create a model of the required services.
\cite{arsanjani2006service}. Once the SOA domain model is finalized, it is
realized by either writing new code, using a third party
implementation, or reusing matching portions of the existing monolithic
implementation and wrapping them into services. For large business systems
that run into millions of lines of code, realizing the services by writing
new code is infeasible. On the other hand, using a third party
implementation is not always possible. While it may be possible to find
third party implementations of very commonly used services such as sales,
locating domain specific services or core business services would be
difficult. Even when a third party service resembling the required service
has been identified, the degree of matching and compatibility with other
proprietary in-house services is an important factor.  Therefore, it is
generally agreed that the key to successful migration to SOA is the ability
to reuse the functionality already implemented in the existing system.

However, given an abstract service description, locating parts of the
source code in the existing system that implements that service is not
easy, and has been a challenge for the SOA research community
\cite{Ricca2009}. There are several reasons for this.  In large monolithic
systems, generally the source code runs into millions of lines of code
making it hard to understand and search. Moreover, knowledgeable developers
might have left exacerbating the problem of locating relevant portions of
code.  Systems documentation is sparse or in some cases non-existent. There
is often a lot of code serving utility purposes (i.e., ``plumbing code''),
that gets in the way of matching core business services to their
implementations.  The most challenging aspect of this problem, however, is
that the terminology used at the level of business process is likely to be
different from the one used at the code level to name files, variables and
functions.

For of all these reasons, manually identifying parts of code that implement
a given service is infeasible for large systems with thousands of files. At
the same time, almost all practitioners we talk to agree that a completely
automated approach is unlikely to yield good results due to the complexity
of the problem.  Therefore, we believe, a semi-automated approach where a
knowledgeable developer follows a clear methodology with tool support is
the way forward. Even here, it is not clear what methodology or heuristics
a developer ought to follow while attempting to identify whether an
abstract service has been implemented in the source code, and if so
how. What would be very helpful is a detailed, real-life case study of the
problem of matching a model of services to an implementation. Such a study
would identify the challenges in this problem, suggest a road-map of
specific technical problems to solve in order to arrive at solutions,
and come up with some initial results towards a solution.  Unfortunately,
to the best of our knowledge, there are no case studies of this nature
reported in the literature.

The goal of this report is to address these issues. We carry out a detailed
case-study to identify (a) the feasibility of and challenges in this
problem, (b) the characteristics of a good match between a model and an
implementation, and (c) features in code and heuristics that are most
suited to (partially) automate a solution to this problem.  In our case
study, we began with a structured list of service descriptions in the ERP
domain provided to us by practitioners, and attempted to match these in the
source code of an open-source Java-based ERP system called JAllInOne. We
first manually located portions of code in JAllInOne that implemented the
given services in the most precise manner possible. We then designed some
semi-automated heuristics that we hypothesized would help partially
automate a solution to the problem, and evaluated them against our (ideal)
manual matching. These set of heuristics, implemented as tools, can assist
a developer in matching service implementations in abstract service descriptions to the existing code.

The contributions of this work are:
\begin{itemize}
\item A detailed manual methodology for mapping a model to an
  implementation. We believe ours is the first approach to match a real
  domain model to a real application to the fullest extent
  possible. Another novelty is the way we use code features to \emph{find}
  matches for services, and then use the GUI to \emph{validate} our
  matches.

\item We describe our experiences during the matching, show representative
  results, highlight challenges, say what works and what does not work, and
  justify the strengths of our proposed methodology.

\item We present several observations about the structure of the
  application we analyze (which is typical of monolithic business
  applications in many ways), that are likely to be useful to researchers
  and practitioners working in this area.

\item We present a set of automated heuristics for model to code matching,
  that could be useful in partially automating the matching problem. We
  validate them against our earlier mentioned manual study, and identify
  which ones among them work well and which ones do not work so well.
  
\end{itemize}

The rest of this report is structured as follows. We describe, the real-life
model as well as the application that we use for our case study in
Sec.~\ref{sec:model-and-app}, goals, challenges and overview of the case study in Sec.~\ref{ssec:casestudy:goals}, and step by step manual methodology used in the case study in detail in
Sec.~\ref{sec:casestudy}. Make some observations that came out of
the case study in Sec.~\ref{sec:observations}. We propose and evaluate
certain heuristics for the matching problem in
Sec.~\ref{sec:heuristics}. We  survey related work in
Sec.~\ref{sec:related-work}, and mention directions for future work in
Sec.~\ref{sec:future-work}. Finally, we conclude in
Section~\ref{sec:conclusions} with a summary of our contributions, as well
as potential takeaways from these for Infosys.

\newcommand{\jallin}{JAllInOne}
\newcommand{\oswing}{OpenSwing}

\section{Description of model and application}
\label{sec:model-and-app}

\subsection{The model}

\begin{figure}
\centering
\includegraphics[width=3.5in]{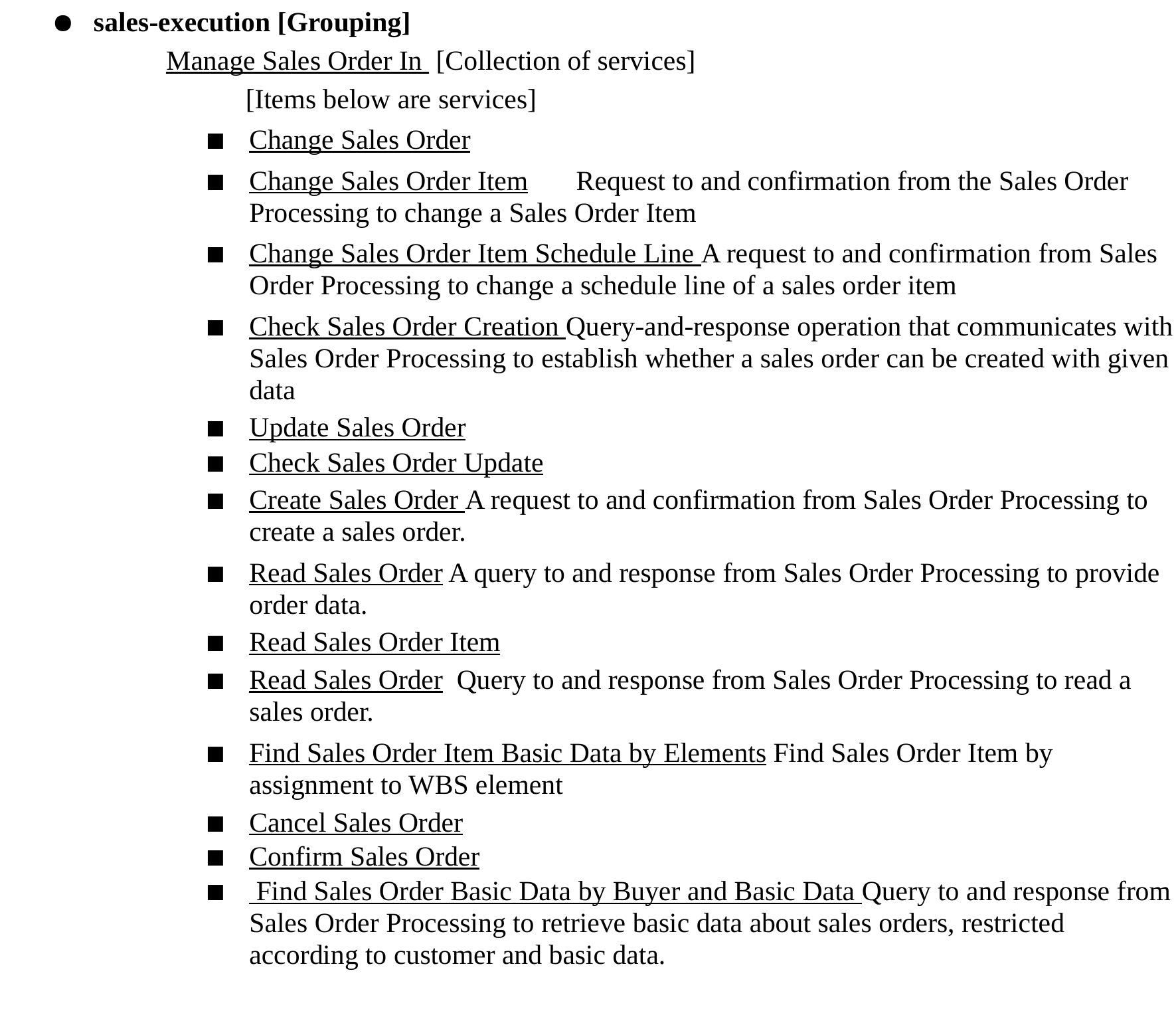}
\caption{A fragment our domain model}
\label{fig:model-screenshot}
\end{figure}

The key artifacts used in our case study were a \emph{model} and an
application. The model was created independently by domain experts in a
major global software services company, and is an English language
description of a representative set of services required in the ERP
(Enterprise Resource Planning) domain. A service is a user-recognizable
high-level functionality. A subset of the model is shown in
Fig.~\ref{fig:model-screenshot}. In the model each service has a name and
a description; often the description is very brief, and does not contain
much information beyond what is implied by the name.  As can be seen,
services are grouped into \emph{collections}, and collections into
\emph{groupings} (there are other collections within the Sales Execution
grouping that are not shown in the figure).  A service collection refers to
a set of services acting on a common \emph{entity}, and differing only in
their \emph{action} on the entity. For instance,
Fig.~\ref{fig:model-screenshot} shows the service collection Manage Sales
Order In (referred to hereafter as ``Sales Order,'') containing several
services that pertain to different actions on
sales orders. It is possible for multiple collections to be based on the
same entity. A grouping is a set of service collections acting on related
entities. For instance, the sales-execution grouping shown in the figure
contains other collections such Manage Customer Returns In and Ordering
Out. Examples of other groupings in the model are Account Management,
Demand Fulfillment, and Demand Planning.
Fig.~\ref{fig:manual-match-stats}, column 2, shows the total number of
groupings, collections, and individual services in the model we use.

\begin{figure}
  \centering
  \begin{scriptsize}
  \begin{tabular}{|l|r|r|r|r|}\hline
              & Model & UI & Model $\leftrightarrow$ UI & Model
                 $\leftrightarrow$ code \\
              &       &    & \multicolumn{1}{|c|}{match}
                & \multicolumn{1}{|c|}{match} \\\hline
  Groupings &   10 & 11 & 4 & - \\
  Collections & 48 & 94 & 9 & 9 \\
  Services &   191 & \~{}256 & 34 & 36\\\hline
  \end{tabular}
  \end{scriptsize}
  \caption{Statistics about the manual matching}
  \label{fig:manual-match-stats}
\end{figure}

As stated in the Introduction, a domain model such as this one plays a very
important role in any exercise in identifying service implementations from
an application. The model helps fix both the exact nature of services
sought, as well as their granularity. Approaches that do not use a domain
model are likely to find services that are too fine grained or too coarse
grained for the requirement in hand. Also, matching against a model makes
it trivial to assign meaningful names to identified service
implementations, which is very useful for the usability of the inferred
services.

The model used in this study is a subset of the full model developed by domain experts (the subset model is summarized in Column 2 of Fig.~\ref{fig:manual-match-stats}), as we were not aware of the full model~\cite{sapmodel} (having 657 collections) at the time of this study. Our part of the model does not capture most of the functionalities of the business domain e.g., services corresponding to customer relationship management, employees, administration, etc.. On the other hand, matching of the subset of the full model, made our study more tractable, and therefore we could do focused analysis of matching service descriptions with their implementation in the application code.

We looked at openly available models also provided by OASIS group (ebxml documents)~\cite{ebxml}. But the ebxml documents contain very coarse grain services, each of which is equivalent to a set of collection. Due to this we could not use the ebxml documents.

\section{The application}
\label{sec:application}

For our study we use the open-source application \jallin~\cite{jallinone},
which is an ERP application designed for medium and small-scale
companies. It is developed using \oswing\ framework~\cite{openswing}. The reasons why we chose \jallin\, apart from the
fact that it was open-source, are that it is in the same domain
as the model (i.e. ERP), is reasonably well-modularized, making the manual
study somewhat more tractable, and is written completely in a single
language -- Java -- making it a good test-candidate to implement
analysis techniques to partially automate the problem of
service mining. \jallin\ has 1089 files (classes), contained in 258
directories and subdirectories, with 223,241 lines of code. The version of JAllInOne (0.9.21) used in the case study can be downloaded from the given link~\cite{jallinone-source-code}.

\begin{figure}
  \centering
  \includegraphics[width=3.0in]{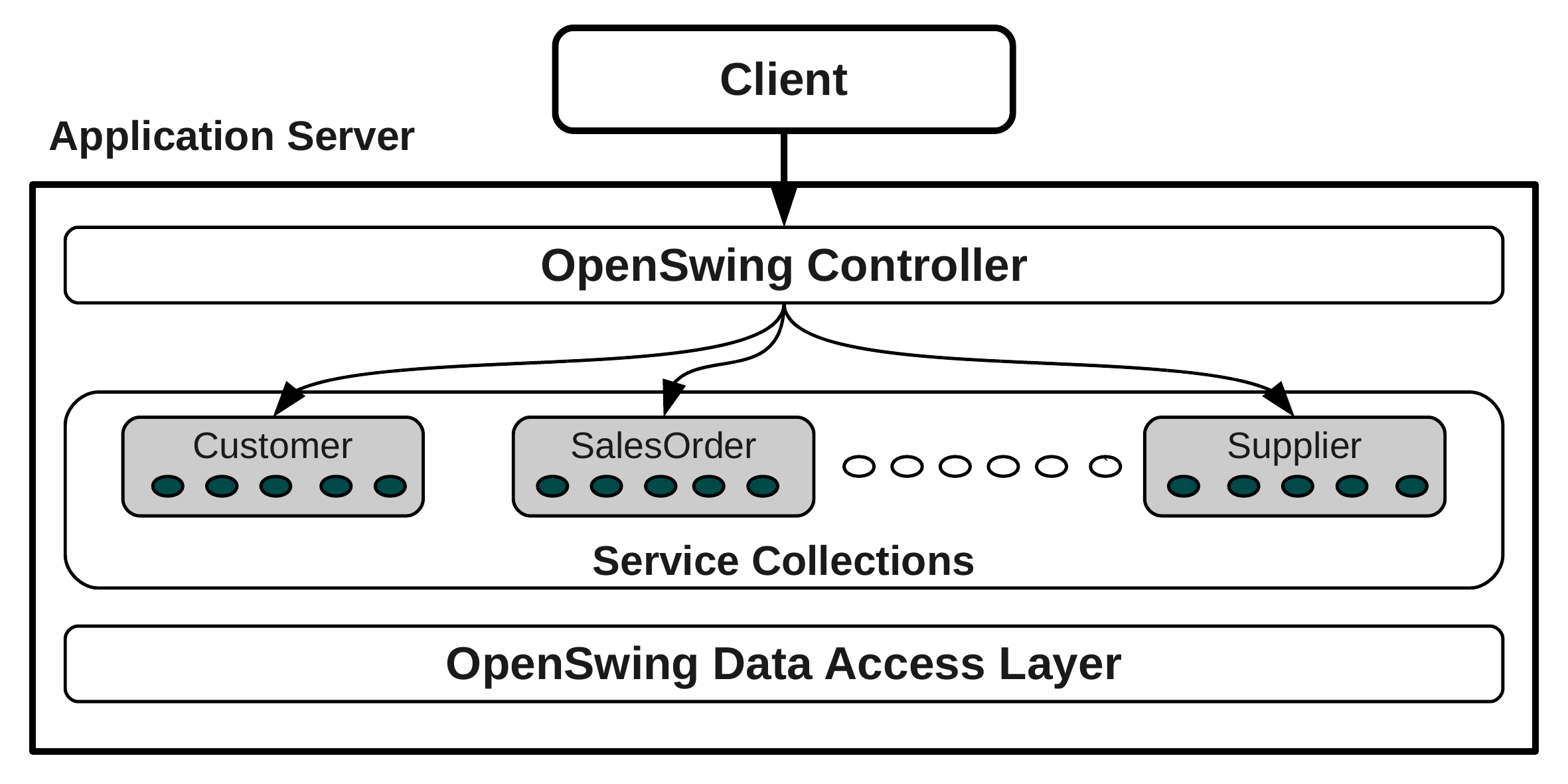}  
  \caption{\jallin\ architecture}
  \label{fig:jallin-architecture}
\end{figure}

The architecture of \jallin\ is depicted in
Fig.~\ref{fig:jallin-architecture}. It consists of a UI client, a
``server'' which contains the classes that provide actual functionality,
and a ``controller'' which receives commands from the client and invokes
appropriate classes in the server. While analyzing the code we restrict our
attention to the server, though our study does involve running the
application and invoking its UI. Hence, whenever we talk about classes or files in
the application, we mean the classes or files in the server.

\begin{figure}
\centering
\includegraphics[width=3in]{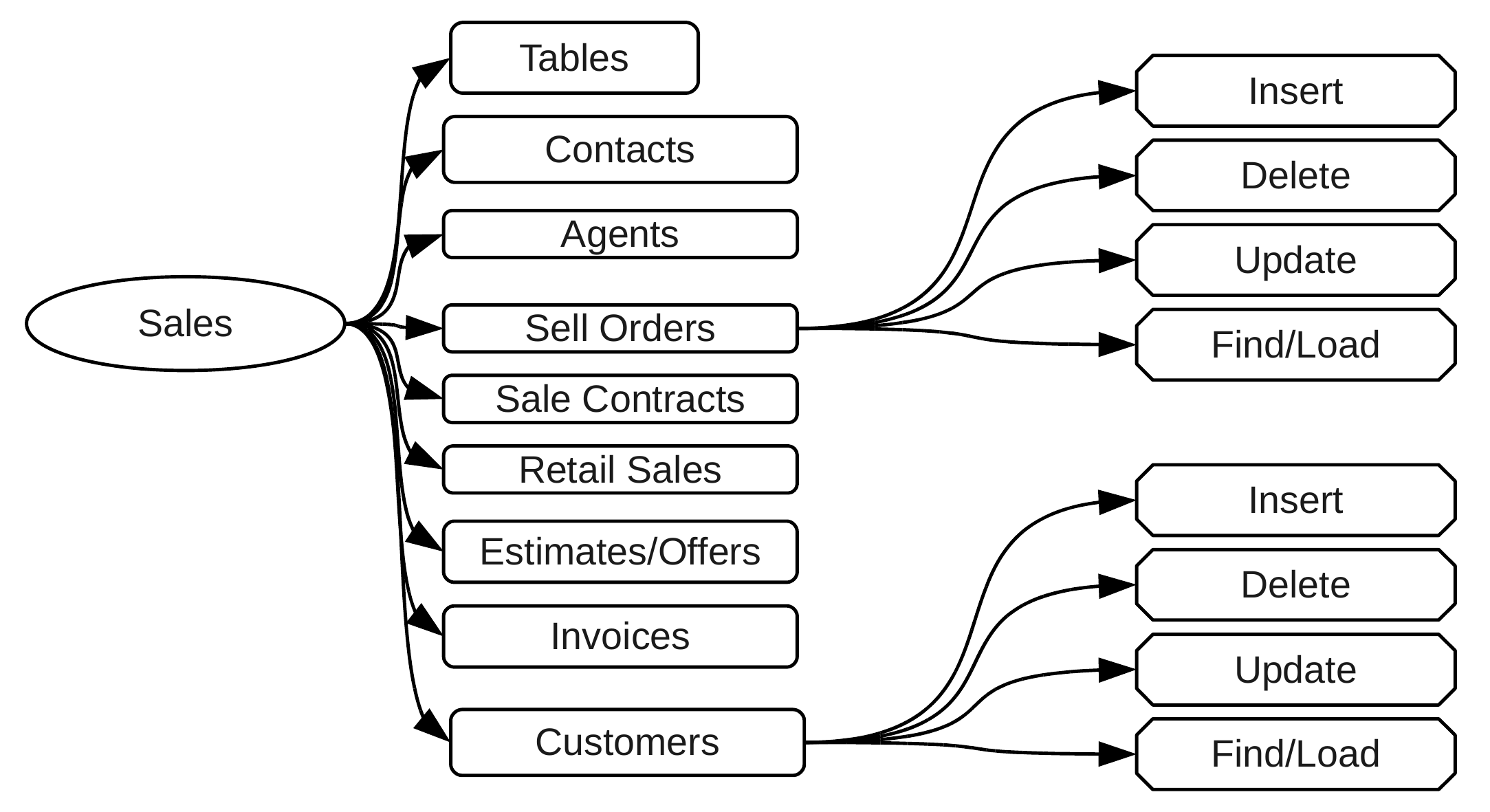}
\caption{Abstract representation of a subset of the \jallin\ UI}
\label{fig:jallin-UI}
\end{figure}

\begin{figure*}
\centering
\includegraphics[width=6.5in]{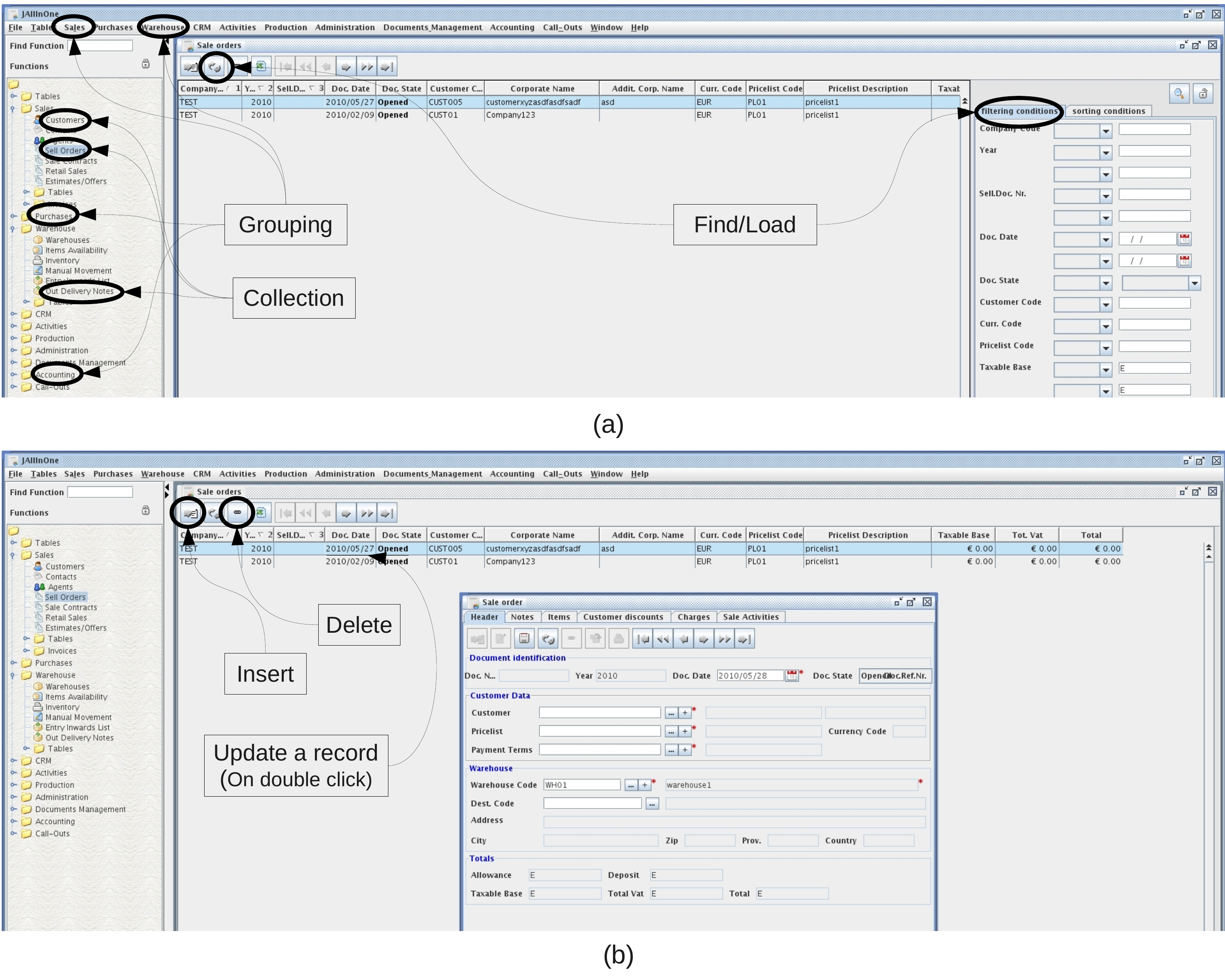}
\caption{Screen-shot of the \jallin\ UI. (a) Grouping-box points to some of the groupings (menus). Some of the collections (menu-items) are pointed to by Collection-box. 
The load action in the UI-abstraction consists of the actions pointed to by Find/Load box. (b) Insert, Delete and Update.. boxes point to the corresponding actions. Sub-window shows the form which appears on clicking the button ``Insert Sell Order'' (button pointed by Insert-box).}
\label{fig:jallin-screenshot}
\end{figure*}

The UI of \jallin\ is interesting, in that it can be abstracted as a three tiered structure, in which tiers correspond to groupings, service collections, and UI-actions (similar to the three tiers in the model).

Next we describe the architecture of the \jallin\ UI and an abstraction of it which is used in this study. Abstraction of UI helps in matching the model to UI at an abstract conceptual level. Once we have abstract representation of the UI, we can simply invoke the UI action corresponding to a model service to collect its execution trace, which we use in validation phase.  We depict
a part of the UI abstraction in Fig.~\ref{fig:jallin-UI}. From left to
right the nodes depict a grouping (i.e., Sales), some collections under
this grouping, and services under two of the collections (Sell Orders and
Customers).
The third column of Fig.~\ref{fig:manual-match-stats} shows statistics about the UI. 
Overall the \jallin\ UI is structured as a set of menus. 

We now describe the process of creating the UI abstraction manually. In the abstraction, the menus are referred as groupings (e.g., the ``Grouping'' block points to some of the menus in the Fig.~\ref{fig:jallin-screenshot}(a)). Every menu item is abstracted as a collection in the UI abstraction. Some collections in the UI are marked ``Collection'' in the Fig.~\ref{fig:jallin-screenshot}(a). When we select a menu item the corresponding page appears, on which we find all the available actions (in the form of buttons) for the collection.
In the abstraction, we consider only the actions that lead to server requests. The actions (i.e., buttons) which do not make any server request use the local information available on the client side. Some of the actions in the abstraction, are pointed by ``Find/Load'', ``Insert'', ``Delete'' and ``Update'', etc., in Fig.~\ref{fig:jallin-screenshot}(a) and Fig.~\ref{fig:jallin-screenshot}(b). The form for Insert Sell Order service is depicted in the sub-window in Fig.~\ref{fig:jallin-screenshot}(b) screen-shot. This form appears on clicking the button (labeled ``Insert'') corresponding to this service.
We ignore all sub-actions within an action. The result of manually abstracting the \jallin\ UI as described above is shown in Fig.~\ref{fig:jallin-UI}.

Note that fortunately the granularity of collections and services in the \jallin\ UI is similar to the granularity of these elements in model. 
Hereafter whenever we refer to the UI, we mean the UI abstraction, as shown in Fig.~\ref{fig:jallin-UI}.

\section{Goals of the study, and challenges faced}
\label{ssec:casestudy:goals}

The goals of our manual case study were to match as many services in the model as possible to the application. We did this task in two steps. First we matched collections in the model to the application, since every service is an action acting upon a business entity corresponding to a collection in the model. For each matching collection, we identified a set of classes in the code, which we call the \emph{collection implementation}, that implements the functionality of the collection. In the second step we matched the services in the matching collection to subsets of files contained in the collection implementation.
  
Our ultimate goals are modularization of system into collection and service implementations, and reuse of the obtained service implementations. Disjointness (in terms of files) is the primary condition for modularization. Otherwise, after modularization, the system may contain copies of a single file in multiple modules.

Our goal was to do this matching as precisely as possible, so we could identify (a) the feasibility of and challenges in this problem, (b) the characteristics of good matches, and (c) features in code and heuristics that were most suited to (partially) automate a solution to this problem. 

Our first thought was to do the matching by (a) matching the services in the model to services in the UI (described in Section~\ref{ssec:model-to-ui}), and (b) executing each service in the UI, recording the classes reached in the execution trace, and assigning these classes to the service that was executed.  We realized however, that there were various difficulties with this approach that made it infeasible, as listed below.

\begin{figure*}
\centering
\includegraphics[width=6.0in]{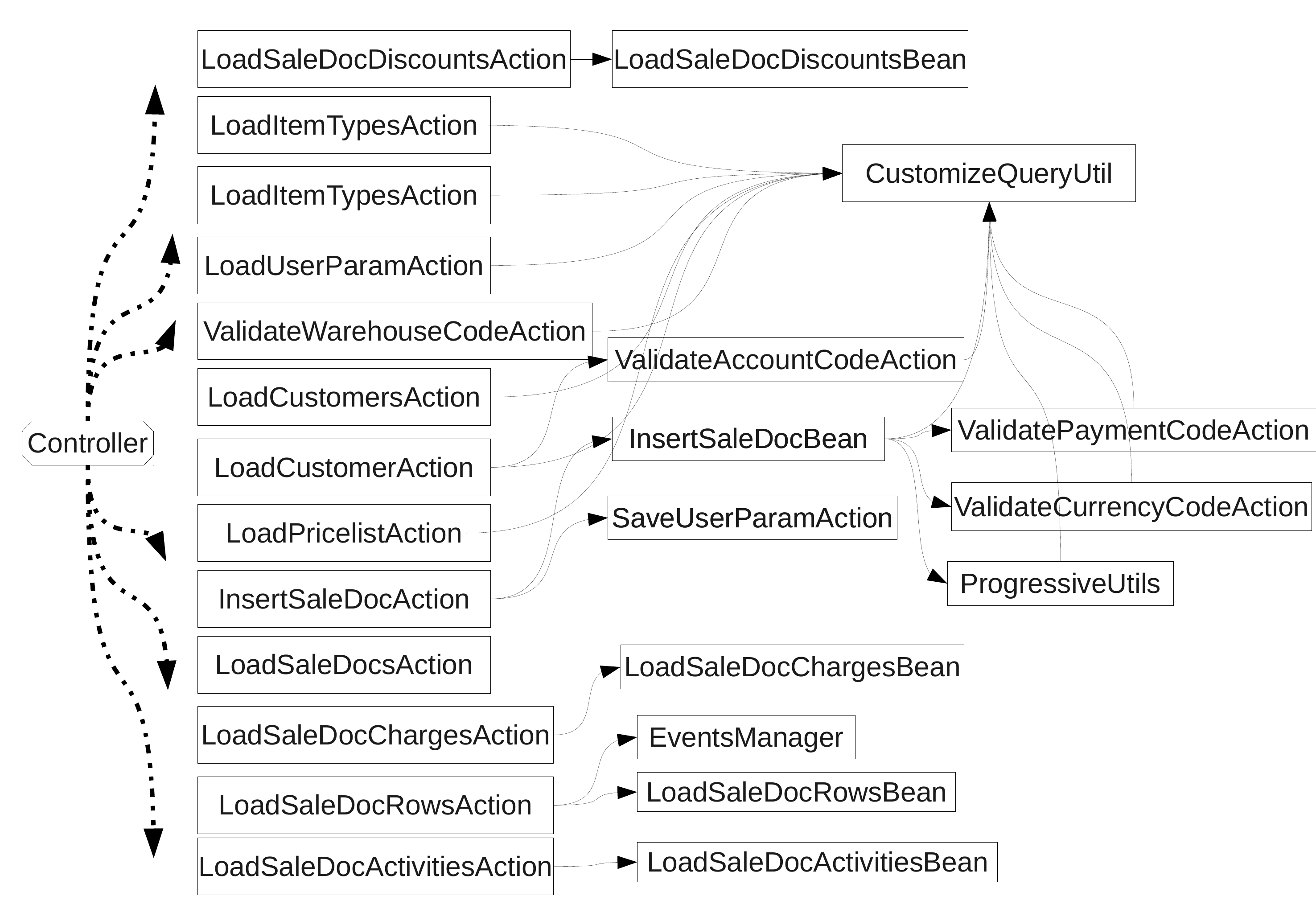}
\caption{Execution trace of Insert Sell Order}
\label{fig:so-ex-trace}
\end{figure*}

\begin{itemize}
\item In many cases the UI does not have sufficient features to allow us to
  guess with confidence whether a certain UI element matches a service in
  the model. For instance there is a collection Vehicle Movement in the
  model. The same label does not appear anywhere in the UI, but the UI
  does have an element Goods Movement. It is hard to ascertain whether the
  two ought to match or not. In other words, we do not have very high
  confidence in our model-to-UI matching. Whereas, in the code there are a
  number of other features (e.g. names of identifiers, comments, names of
  files and directories) that increase the confidence of our matching.

\item Although we earlier mentioned that the UI is organized, like the
  model, in terms of groupings, collections, and actions, in fact many of
  the leaf elements of the UI are not simple services, but compositions of
  services (i.e., mini business processes). For instance, when we execute
  the action Insert under ``Sell Orders'' in Fig.~\ref{fig:jallin-UI},
  which matches to the service ``Create Sales Order'' in the model (see
  Fig.~\ref{fig:model-screenshot}), the UI makes us select a customer
  from a list of customers, an item from a list of items, etc., in order to
  populate the sales order. The fact that each UI service is actually a mini process, is confirmed by the trace of classes that
  are reached during this execution, as shown in
  Fig.~\ref{fig:so-ex-trace}.  The classes in the first column (after the
  Controller) get invoked directly by the Controller, whereas the other
  classes are invoked transitively (through a chain of invocations).  In our understanding only the classes
  InsertSaleDocAction (called by the Controller) and InsertSaleDocBean
  (called by InsertSaleDocAction) constitute the implementation of Create
  Sales Order proper.  Classes such as LoadCustomersAction and
  LoadItemTypesAction are executed as part of the composite UI action
  Insert Sell Orders, but in fact match other services in the model, and
  should not be included in the implementation of Create Sales Order.  It
  would be quite difficult to make such decisions correctly just from the
  information in the trace.

\item The UI does not directly expose to the user all the services that
  exist both in the model and in the code. For instance, Confirm Sales
  Order and Validate Sales Order are services in the model that have a
  matching implementation in the code, but do not match any action in the
  UI by name. The classes corresponding to these services are invoked
  implicitly as part of other composite actions in the UI; therefore, it
  would be difficult to match these classes to their respective matching
  services using just the information in the trace.

\item Several of the classes in Fig.~\ref{fig:so-ex-trace} are utility
  classes, with no business logic (e.g. EventsManager). These ought not be
  included in the implementation of any collection. It would be hard to
  identify and separate out such utility classes from just the traces.

\item Not all applications have a UI; for example batch processing systems which are used in the back-end of several existing systems. We would like our case study methodology and results to also extend to applications without
  UIs. Hence we choose to use the UI, when available, for validation of a
  model-to-code match, and to not require a UI to be present to do the
  matching.
\end{itemize}

\begin{figure}
\centering
\includegraphics[width=3.5in]{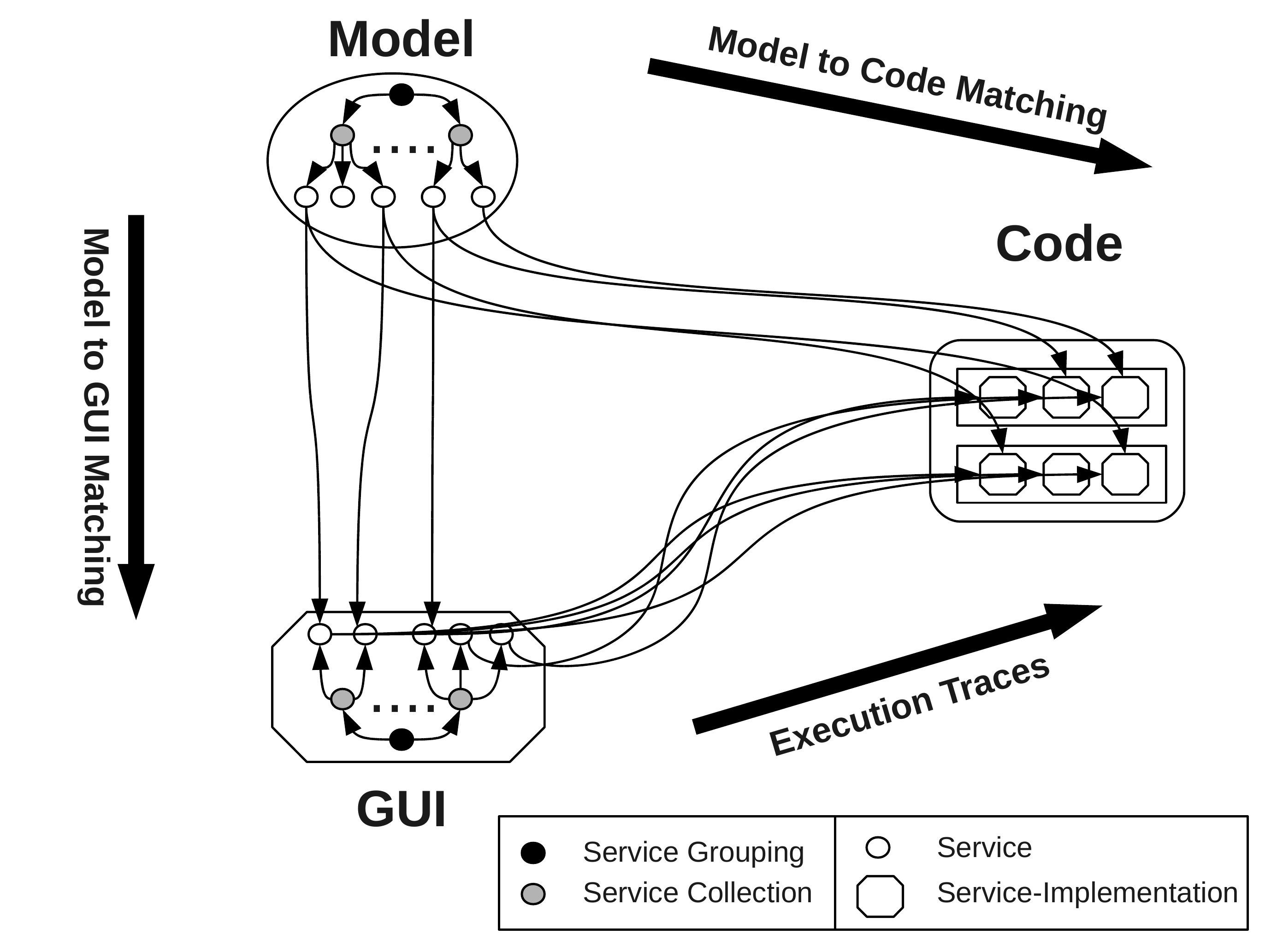}
\caption{Three step approach to matching model with application.}
\label{fig:threelegs}
\end{figure}

For these reasons we decided to follow the approach of independently
matching the model to the code (as described in
Section~\ref{section:model-to-code}). After matching the model to the code,
we used execution traces to validate this matching
(Section~\ref{ssec:validation}). This entire process is summarized
in Fig.~\ref{fig:threelegs}.

The model to code matching presented its
own set of issues, namely (a) large size of the application, (b)
ambiguities in deciding the boundaries between the implementations of
different collections/services, (c) the use of terminology  in the
identifiers and comments in the application that is different from that in
the model, e.g., actions (insert vs. create, load vs. ``read''$\mid$``find'', etc.), terms present in the description of services, some collection names  (e.g., item vs. material), and (d) the presence of a large number of non-business-logic
related (utility) classes. We describe in
Section~\ref{section:model-to-code} how we overcame these difficulties.
One difficulty we did \emph{not} face was interleaving of code
corresponding to different services or collections within a single
class. \jallin\ is well-modularized in this respect, whereas older
applications in legacy languages are often
not~\cite{rugaber-interleaving}. Since our focus was on understanding the
issues and difficulties listed above, it helped us that our application did
not present to us the additional difficulty of interleaving.

\section{Steps in Manual Methodology}
\label{sec:casestudy}

\subsection*{Step 1: Matching model to code}
\label{section:model-to-code}

The goal of this part of the case study, as mentioned in Chapter~\ref{ssec:casestudy:goals}, is to match each collection (and within that, each service) in the model with the code, and associate a set of classes (or files) with each collection (and service) that has a match in the code. We call the set of classes associated with a collection (service) the \emph{implementation} of the collection (service). We wished to do this task as precisely as possible, using all the features available in the code, as well as our human intelligence, so that the results of the study would form a basis for devising and evaluating (partially) automated techniques. We first devised a set of necessary (i.e., minimum) guarantees that the manual matching had to satisfy:

\begin{itemize}
\item The implementations of the collections are non-overlapping (as sets of classes), and the service implementations within a collection implementation are also non-overlapping. We have already described the reason for this in Chapter~\ref{ssec:casestudy:goals}.
\item If a class $A$ had only one calling class $B$ then $A$ is in the same collection as $B$.
\item If a class $A$ was calling only one class $B$ then $B$ is in the same collection as $A$.
\end{itemize}

The methodology we followed to match a collection $C$ was as follows. Firstly, identify a set of \emph{seed} files for the collection using features in the files that strongly associate it with the collection. 

Once we had seed files for collection $C$, we followed call-graph edges from these files, in both directions, looking to add neighboring files to the collection. We call the process ``expansion'', and it terminates when no more files can be associated with collections. We verified at the end of the process to the best of our ability that the files associated with each collection indeed constitute the implementation of the collection. 

We now present the details of the approach summarized above. Firstly, we enumerate certain properties of files that we use in seed-finding and in expansion.

\begin{itemize}
\item $P_1$: Accesses one or more database tables ($T_{C}$) pertinent to the collection $C$.
\item $P_2$: Accesses majority of the fields (attributes) of one or more tables in $T_{C}$.
\item $P_3$: Name of the file has some similarity with the collection name or a table name in $T_{C}$.
\item $P_4$: Contains comments indicating its relevance to the collection $C$.
\item $P_5$: Most of the callers and callees feature most of the above properties; i.e. the file is surrounded by other files that belong to the implementation of collection $C$.
\item $P_6$: The file is located in the directory where most of other files of the specific collection are located.
\end{itemize}

A set of rules $R_S$ that a file should satisfy to be a seed file of the collection $C$, is given below.

\begin{itemize}
\item $S_1$: The file exhibits property $P_1$, and
\item $S_2$: The file conforms to the majority of the other five properties ($P_2-P_6$), and
\item $S_3$: The file shows closer proximity (based on $S_1$ and $S_2$) to the given collection $C$ than other collections.
\end{itemize}

Similarly in expansion phase, a file was added to the implementation of the collection $C$ if either rules $E_1$, $E_2$ and $E_3$; or rules $E_2$ and $E_4$; or all four rules; were satisfied:

\begin{itemize}
\item $E_1$: Satisfies some of the properties among $P_1-P_6$. 
\item $E_2$: Has close proximity (i.e., called by or calling) to a seed-file or any other file already assigned to the collection implementation.
\item $E_3$: Shows closer proximity (based on $E_1$ and $E_2$) to the given collection than other collections.
\item $E_4$: Is not called by any of the files of any of the other collections.
\end{itemize}

\begin{figure*}%
 \small{
\begin{center}
\begin{tabular}{lllllllllllll}\hline\\
{$Collection$ $Name:  $}× & {$|T_{\mathit{UA}}|$} × & {$|T_{\mathit{AM}}|$} × & {$|T_{C}|$} × & {$N_{P_1}$}× & {$N_{P_2}$} × & {$N_{P_3}$} × & {$N_{P_4}$} × & {$N_{P_5}$} × & {$N_{P_6}$} × & {$N_S$} × & {$N_E$}× & {$|S_C|$} × \\ \hline\\ 
\\
Customer	× & 1× & 1× & 1× & 14× & 17× & 28× & 6× & 0× & 6× & 6× & 0× & 6× \\ 
\\
Inbound Delivery× & 1× & 1× & 2× & 20× & 11× & 12× & 8× & 4× & 23× & 8× & 1× & 9× \\
\\
Material 	× & 1× & 4× & 2× & 39× & 11× & 52× & 13× & 1× & 13× & 7× & 2× & 9× \\ 
\\
Outbound Delivery× & 1× & 1× & 2× & 20× & 14× & 12× & 8× & 4× & 23× & 8× & 1× & 9× \\
\\
ProductionOrder × & 1× & 4× & 5× & 12× & 16× & 12× & 14× & 5× & 14× & 10× & 4× & 14× \\
\\
Purchase Order  × & 1× & 1× & 2× & 19× & 23× & 22× & 14× & 8× & 23× & 12× & 5× & 17× \\ 
\\
Sales Order 	× & 1× & 4× & 2× & 31× & 20× & 65× & 19× & 8× & 32× & 14× & 7× & 21× \\ 
\\
Sales PriceList	× & 2× & 4× & 2× & 17× & 20× & 24× & 6× & 0× & 11× & 10× & 0× & 10× \\ 
\\
Supplier	× & 1× & 1× & 1× & 14× & 39× & 26× & 6× & 0× & 6× & 6× & 0× & 6× \\ \\ \hline
\end{tabular}
\end{center}
}
\vspace*{2pt}
\caption{Statistics of the model to code matching. $T_{\mathit{UA}}$,
  $T_{\mathit{AM}}$ and $T_{C}$ refer to the number of unambiguous tables,
  ambiguous tables, and tables (after ambiguity resolution), for each
  collection.  $N_{P_1} - N_{P_6}$ are the number of files satisfying properties $P_1 - P_6$ resp.. $N_S$, $N_E$ and $|S_C|$ denote the number of seed files, expanding files and total files for the given collection. }
\label{fig:collection-features}
\end{figure*}

The summary of the results of allocating files to collections using the rules mentioned above is shown in Fig.~\ref{fig:collection-features}. We manually determined the set of database tables $T_C$ for each collection $C$ at the start of the case study. In the figure, $T_{UA}$ and $T_{AM}$ refers to the tables which can be associated to the collection unambiguously and ambiguously, respectively. The tables finally associated with each collection, denoted by $T_C$, is got by resolving some of the ambiguities in $T_{AM}$ and adding it to $T_{UA}$. This is how we disambiguated the tables $T_{AM}$: among the files accessing an ambiguous table, if any of the files also accessed any unambiguous table or if the comments inside the file seemed to indicate its relevance to the collection $C$, then we associated this table with $C$. If an ambiguous table was accessed by a large number of files or if the files did not give sufficient confidence the table was left out of $T_C$.

Columns $N_{P_1} - N_{P_6}$ summarize the number of files satisfying properties $P_1 - P_6$ for each collection. The number of files we marked as seed files and files in expansion are given in the columns $N_S$ and $N_E$ respectively. The last column $|S_C|$ gives the number of files in the implementing set of files ($S_C$) associated with each collection ($|S_C| = N_S + N_E$). For each collection, the file names in the implementing set ($S_C$) are given in the second column of Fig.~\ref{fig:allfiles}.

As we observe in rules $S_2$, $S_3$, $E_1$ and $E_3$, we often took subjective decisions using our intuition to include a file in a collection.

Once each collection was assigned a set of files, we partitioned this set of files among the services in the collection. This is often easy to do; the seed files of a service often contain in their name or in the names of identifiers within them the \emph{action} word associated with the service (e.g. load, create, delete). Once the seed files of each service within a collection have been identified in this way, we expand from the seeds and hence partition the set of files associated with the collection among its services using a process similar to the one described above for associating files with collections. 

The fifth column in Fig.~\ref{fig:manual-match-stats} summarizes the information about the number of model elements that found matches in the code.

\begin{figure*}
\centering
\includegraphics[trim=0 0 0 0,clip=true,width=6.5in]{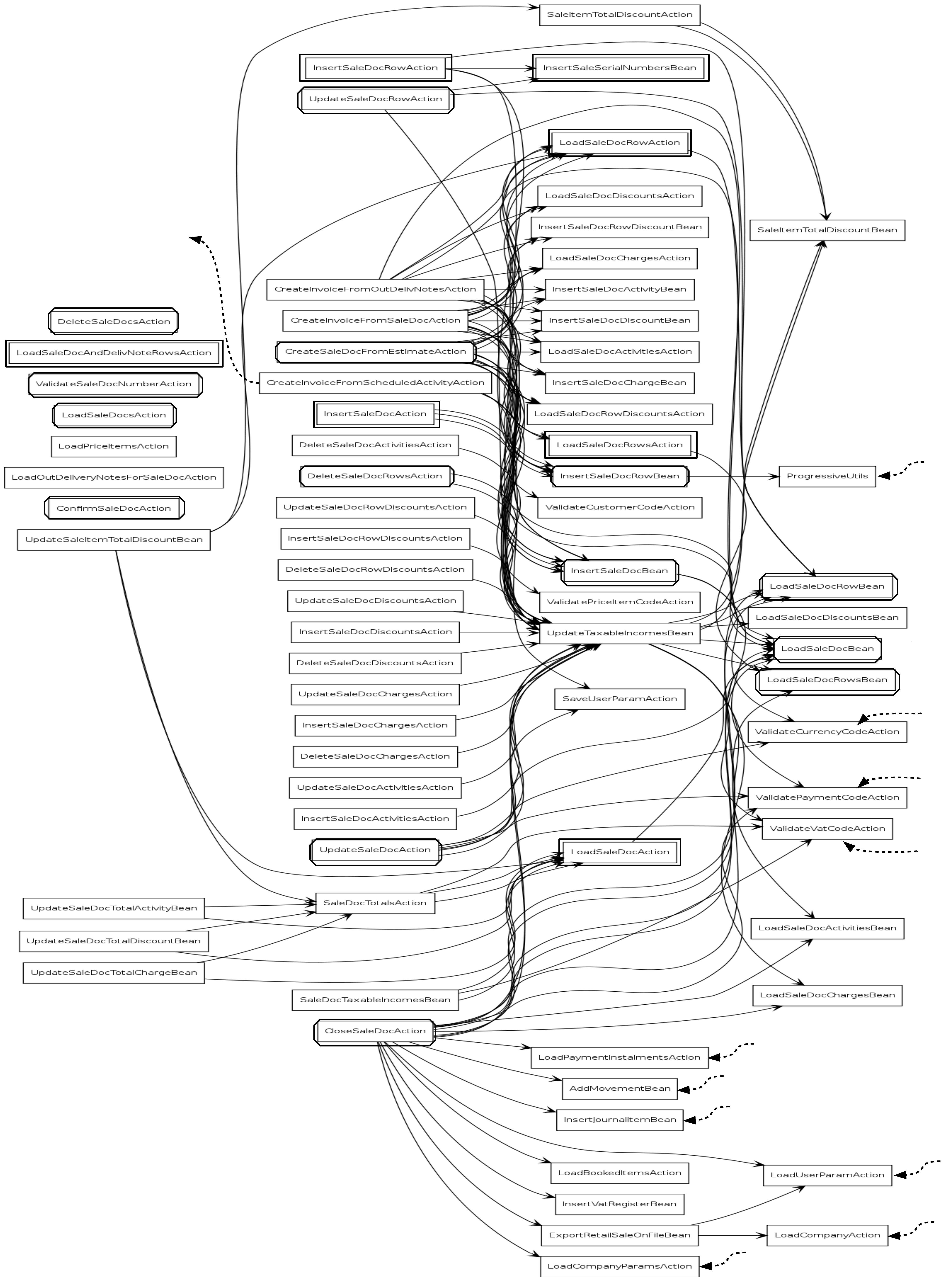}
\caption{Call graph for files of sales-order collection and other files in surrounding (callers, callees and files in the same directory). Octagonal-boxes and double-boxes refer to the seed-files and files found in expansion, respectively.}
\label{fig:socg}
\end{figure*}

We now give the details of various tools used in this step. We also use the tools used in this step during ``Validation'' (Section~\ref{ssec:validation}) and partial automation of the model to code matching (Section~\ref{ssec:semi-automated-approach}). We built a tool over three existing tools ``DoxyGen''~\cite{doxygen}, ``Graph-Easy''~\cite{grapheasy} and ``GraphViz''~\cite{graphviz}. ``Doxygen'' is used to generate the call graph for the application code. It generates call graphs in pieces, i.e., a graph for each function, containing callers and callees of this function only. We first parsed the .dot files in the output of ``Doxygen'' to get the full call-graph from the smaller call graphs. We then constructed an abstracted call graph, in which every file is taken as a node and the edges are among files only; i.e., we abstracted all functions in a file as single node. An edge is added in the call graph from file $f_1$ to $f_2$, if a call edge is found in the original call graph emerging from some function in $f_1$ and is incident on some function in $f_2$. Next we remove all nodes corresponding to and edges adjacent to any file that is not a server file.

\subsubsection*{Illustration of model to code matching}

As an example, we depict in Fig.~\ref{fig:socg} the files we matched with the Sales Order collection in the model (see Fig.~\ref{fig:model-screenshot}), along with the files which are in the same directory as the matching files, and the immediate callers and callees of all these files. Each node is a file in the implementation, and the edges are abstracted call edges. The files without any call-edges incident are the entry points. The dashed call-edges are ones whose other end is outside. Each dashed call-edge in fact represents one ore more call-edges with the same direction. The octagonal boxes are the seed files and the double boxes are the files obtained by \emph{expansion} for the Sales Order collection. All other files depicted in the figure are ones that were not included in the Sales Order collection implementation. The following examples describe a few representative cases in seed finding and expansion process for the call graph shown in Fig.~\ref{fig:socg}.

We first determined that among the 119 database tables used in the application DOC01\_SELLING and DOC02\_SELLING\_ITEMS are the tables most pertinent to the Sales Order collection  (see Fig.~\ref{fig:collection-table} for a list of other matching collections, and their pertinent tables); i.e., these constitute the set $T_{SalesOrder}$. Based on this determination we identified the seed files of the collection satisfying the rules $S_1 - S_3$. 
Based on these rules most of the files were easy to identify as seed files (unambiguously), but a file ``CloseSaleDocAction'' was an ambiguous seed file for Sales Order collection.
Next, we discuss the logic used to disambiguate the ambiguous files. We do not discuss about unambiguous files, as it was obvious to either discard or accept an unambiguous file in the implementing files set based on the given rule sets .

The file ``CloseSaleDocAction'' was strong in all features except $P_5$. Based on the name and comment inside (``Action class used to close a sale document...'') we considered it a strong candidate for being a seed. It satisfied all the three rules ($S_1 - S_3$) for a seed. It however did not show very high confidence for $S_3$. This shows that the decision of assigning a class to a collection in the case when all features are not satisfied is a highly human dependent decision.

We now discuss expansion, in which we sought for the files satisfying the rules given for expansion (i.e., $E_1$, $E_2$ and $E_3$; or $E_2$ and $E_4$; or all $E_1-E_4$). In the majority of cases non-seed files could be unambiguously placed in the Sales Order collection based on the given rule set. Several of the non-seed files that we have assigned to this collection are in it even though they are connected to files outside the collection. This was due to the preponderance of evidence to assign them to Sales Order. 
A lot of files not in the collection also have ``sale'' and ``doc'' in their name. Most of the files also are in the same directory as all other files in this collection. 

An interesting case we noticed was the file ``InsertSaleSerialNumbersBean''. Although it did not satisfy most of the given features, it was called by only two  files  (``UpdateSaleDocRowAction'' and ``InsertSalesDocRowAction''), both of which had strong correspondence with the Sales Order collection. Therefore this file also was added to the Sales Order implementation.

Several files were located in the same directory as the files in the collection, but did not access any of the tables pertinent to Sales Order collection (e.g., ``InsertSaleDocDiscountsAction'', ``DeleteSaleDocChargesAction'', ``ValidateVatCodeAction'', ``AddMovementBean'', etc.). The comments inside these files also do not give any confidence towards their relevance to the collection. Therefore, these files do not satisfy the rules for being a seed file or an expanding file , and we could prune these files unambiguously. 

In some cases, like in ``UpdateTaxableIncomeBean'', although the name of the file gave no indication that it ought to belong to this collection, upon closer perusal, we determined that the file (a) accessed the table DOC02\_SELLING\_ITEMS, and (b) had the comment ``Description: Help class used to update all taxable incomes for all items and activities $\ldots$ for the specified sale document $\ldots$''. Overall this file showed good relevance to Sales Order collection. But this file was accessing tables other than Sales Order tables and was called by and calling several files outside the Sales Order implementation (i.e., showed good relevance to files outside the Sales Order implementation). Due to all these reasons we did not include this file in the implementation.

\subsection*{Step 2: Matching model to UI}
\label{ssec:model-to-ui}

The goal of matching the model to the UI is to validate the model-to-code matching at finer level. We use the abstraction of UI illustrated in Fig.~\ref{fig:jallin-UI} for matching the model to the UI.

\begin{figure}
  \begin{scriptsize}
  \begin{tabular}{p{1.6in}p{1.6in}}
    \multicolumn{1}{c}{\underline{Model}} &
    \multicolumn{1}{c}{\underline{UI}} \\

    \textbf{Purchasing and Sourcing (A):} Supply management (2) &
    \textbf{Purchases (A):} New invoice from 
    Delivery notes (8), Buying orders (3), Suppliers (2)\\

    \textbf{Account Management (B):} Price management (4), Customer (5) &
    \textbf{Accounting (B):} \\

    \textbf{Sales Execution (C):} Sales order (6) & \textbf{Sales (C):}
    Sell orders (6),
    Customers (5), Sale price list (4)\\

    \textbf{Warehousing and Storage (D):} & \textbf{Warehouse (D):} Out
    delivery notes (9)\\

    \textbf{Supply Planning:} Material (1) & \textbf{Table:} Items (1) \\

    \textbf{Procure to pay:} Purchase order (3) & \textbf{Production:}
    Production orders (7)\\

    \textbf{Demand Fulfillment:} Production order (7), Inbound delivery
    notes (8),
    Outbound delivery notes (9)& \\
  \end{tabular}
  \end{scriptsize}
\vspace{5pt}
  \caption{Summary of model to UI matching. Each entry is of the form
    \textbf{Group:} \emph{collections}. Groups/collections with the same
    label match.}
  \label{fig:model-to-ui}
\end{figure}

We summarize the results of the model-to-UI matching process in
Fig..~\ref{fig:manual-match-stats}, Column~4, as well as
Fig.~\ref{fig:model-to-ui}. We first tried to match groupings in the
model with groupings in the UI. From the ten groupings in the model and
eleven in the UI (see Fig.~\ref{fig:manual-match-stats}) four pairs were
matching. These four are shown in the first four rows in
Fig.~\ref{fig:model-to-ui} labeled A-D. For brevity, we show matching groups and collections only in the figure. Note that the matching groups do not have identical names; we had to use our intuition and guess
that they match.

Next, we checked if the collections within the four matching groups matched each other. We found that only two such pairs of matching collections (under matching groups) existed; see Collection~2 under Group~A and
Collection~6 under Group~C match. The other collections under the matching
groups did not match each other. Hence, our finding is that the groups in
the UI basically did not align too well with the groups in the model.

Next we exhaustively tried to match all remaining collections in the model
(46 of them) with all remaining collections in the UI (94 of them), and
found that the seven pairs of collections labeled as 1, 3-5, 7-9 matched. In other words,
we found a total of 9 pairs of matching collections between the model and
UI. These are labeled~1 through~9 in Fig.~\ref{fig:model-to-ui}. Note
that as with groups, we had to use domain knowledge, synonyms, as well as
our intuition to find matching collections that did not have identical
names. Within the 9 matching collections there were~46 (resp.,
approximately~36) individual services in the model (resp., UI). Of these~46
services in the model, 34 had matching services in the UI. Note that the
model to UI matching was many-to-many; sometimes multiple model services
matched a single UI element, and sometimes multiple UI elements matched a
model service. However, most of matches were one-to-one.

Note that most of the collections in the model and UI in fact did
\emph{not} match each other. This was due to following reasons. The domain model was meant
for large companies, and had groupings such as Product Development, Supply
Planning, and Demand Planning, which had no match in JAllInOne, which is
meant for small enterprises. At the same time we were using only
a subset of the model, in which were missing key groupings like CRM and
Production, which are present in \jallin. This said, the matching exercise
still gave us a lot of insight into the challenges involved in it, even
though the actual number of matches was small as a proportion of the total
model/application.

\subsection*{Step 3: Validation of match using execution traces}
\label{ssec:validation}

As discussed in Chapter~\ref{ssec:casestudy:goals}, and depicted in
Fig.~\ref{fig:threelegs}, the goal of this step is to use the model-to-UI
matching we obtained (see Section~\ref{ssec:model-to-ui}, Step~2) to validate the
model-to-code matching (Section~\ref{section:model-to-code}, Step~1).  For this purpose we first collect execution traces corresponding to all matching UI actions. Then based on the execution traces we decide, whether for a matching collection validation passes or fails. Before going into details of validation, we describe briefly the process and the tool used to collect execution traces.

To collect traces by invoking the UI services, we used a tool JIP (Java Interactive Profiler)~\cite{jip}. JIP interacts with the portion of the application on server side through a user defined port. Whenever a request is made to server, it collects the trace of all invoked functions while the request is being serviced by server. JIP generates the output trace in the form of a text file and an XML file. 

It should be noted that the profiler collects only traces of server side functions. The traces do not contain any function invoked on client side. Since all functions on the server side are invoked by controller, we find the controller also contained in every execution trace, which we remove. We call a function (in our abstraction, we use the file containing the function) an entry point in a trace if no other function is calling it in the call sequence of the trace after removal of the ``controller''.
We define an entry point in the application code as a file which is not called by any other file in the call-graph constructed statically (see Sec.~\ref{section:model-to-code}, Step~1)

We next explain the validation criteria for a collection. For any given collection $C$, let $G$ be the set of actions in the UI that have matched the services inside $C$ (as per Section~\ref{ssec:model-to-ui}, Step~2). We execute the actions in $G$ and save the resulting set of execution traces $E$. We say that the validation for collection $C$ has passed if all \emph{entry points} in the implementation of $C$ (as defined above) are reached in the set of traces $E$. 
Intuitively, the validation determines if the entry points in the
collection's implementation have been identified \emph{precisely}, in the
sense that each one identified is indeed an entry point (is reachable
during execution from some UI action that matches some service in the
collection). We do \emph{not} validate \emph{recall}, in the sense that we
are not sure if all classes that ought to be entry points of a collection
are indeed assigned to the collection.  The reason for this is as follows.  If every entry point reached in the
 set of traces $E$ corresponding to collection $C$ was part of the
 implementation of \emph{some} collection (not necessarily $C$) then
 clearly recall for $C$ is 100\%. However, if some entry point $e$ in $E$
 does not belong to any collection's implementation it does not
 necessarily imply that $e$ ought to have been included in collection
 $C$'s implementation. As was discussed in 
 Section~\ref{ssec:casestudy:goals}, the actions in $G$ could be composite
 UI actions, that invoke classes outside the implementation of $C$.
 Therefore $e$ could pertain to some collection other than $C$, but could
 remain unassociated with any collection because in this case study we
 matched only a subset of  the full model with the application.  Therefore
 given our methodology it was not possible to validate recall.

The results of the validation were such that of the 9 collections in the
model that have matches (see Fig.~\ref{fig:model-to-ui}), 7 of them
passed the validation, except Sales Order and Purchase Order (see the discrepancy in the number of services in the model that
matched with the UI, versus those that matched with the code, in
Fig.~\ref{fig:model-to-ui}. The
reason for this is interesting to note. The service Confirm Sales
Order in the Sales Order collection 
(see the second from last service in Fig.~\ref{fig:model-screenshot})
matches a set of files in the code, but its entry point (see class ConfirmSaleDocAction in the leftmost column of Fig.~\ref{fig:socg}) does not show up in the
trace corresponding to the execution of any UI action that matched any collection. The interesting thing to note here is that careful
investigation of validation failures can either reflect
the incompleteness of the UI wrt the model (as in this case), or rectify problems with the
model-to-code matching or in the model-to UI matching. We notice that in case of Purchase Order collection also, validation fails for the same reason.

Note that while we validated the precision of all entry points in each
collection's implementation, we did not validate the precision of the other
classes (i.e. non entry points) included in the implementation. This is
straightforward to do conceptually (following a similar process to the one
described above), but was impractical to validate automatically due to the following reasons: (a) UI is incomplete, i.e., a file f may very well belong to a collection C, but may still be unreachable in any trace generated from UI actions that match C, (b) it is difficult to manually force the execution using UI through all possible paths.

 \section{Observations}
\label{sec:observations}

In this chapter we highlight the main observations and lessons
learnt from our case study.
Some of these observations will form the basis of a heuristic
matching technique we propose in the next section.
The remaining observations are mentioned in the hope that they
could be exploited in future efforts.

\subsection{On the adequacy of UI}

As observed in detail in Chapter~\ref{ssec:casestudy:goals}
a UI cannot be a substitute for a model for various reasons.
However we found the UI useful in several ways:
\begin{itemize}
\item
As described in detail in Section~\ref{ssec:model-to-ui}, by trying
to match the groups, collections, and services in the model to
corresponding elements in the UI, we obtain a quick, rough
estimate of services whose implementations exist in the
application.

\item
The UI is useful for validating the output of a matching
exercise.
%
%
%

%
%
%
%
\end{itemize}

\subsection{Natural tiering of services}

The source code modules in the application implementing various services can be naturally classified into
four ``tiers'' as described below.

\begin{enumerate}
\item
\emph{Top-level services.}
These are services listed in the
model whose implementing files are not called by implementations of other services.

\item
\emph{Middle-level services.}
These are services listed in the
model whose implementing files are called or used by at least one other service implementation in the
application.

\item
\emph{Bottom-level services.}
These correspond to a cluster of source files whose implementation
does not match any service description listed in the
model, but which also contain business logic and either have an
independent entry point or is called by two or more service implementations in the
application.

\item
\emph{Utility services}
A cluster of source files that does not contain any
business logic and is called by two or more services.
\end{enumerate}

\begin{figure*} %
\centering
\includegraphics[width=6.2in]{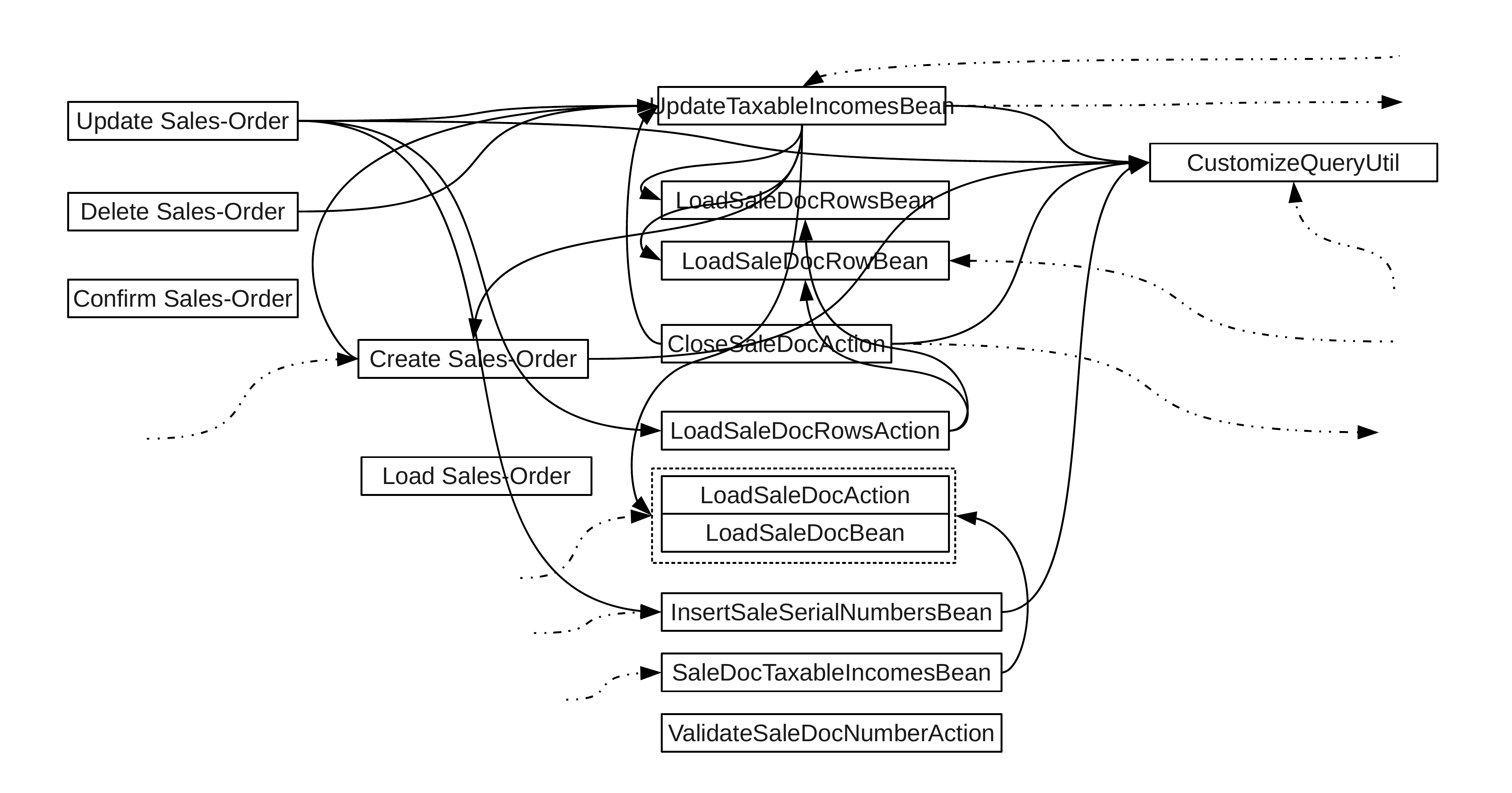}
\caption{4-tier structure of services in Sales Order collection. From left to right, the columns are top, middle, bottom, and utility services respectively.}
\label{fig:sotier}
\end{figure*}

Fig.~\ref{fig:sotier} shows some services in the Sales Order
collection classified according to these tiers.
The four columns from left to right correspond to the top, middle,
bottom, and utility levels respectively.

\subsubsection*{Benefits of four tier architecture}

This tiering structure is important as it proposes a set of natural,
coarse-grained services that are implemented in the application
and that are good candidates for adding to the model, 
namely the bottom-level services.

There are other tiering structures presented in the literature, as in work of Li and Tahvildari~\cite{Li2006}. Their tiering architecture is purely based on the structural organizations (graph transformation and entry points), whereas the proposed tiering architecture is based on the business concepts inherited in the programs. Therefore, our tiering architecture is more closer to the natural understanding of the business softwares.

\subsection{How the model can be enriched}
\begin{enumerate}
\item 
It would be useful if domain experts can provide synonyms for
the terms used in the model.
This makes the task of locating services easier for a human, as
well as allows the possibility of automation.
As an example, we had difficulty understanding the grouping ``procure to
pay'' in the model. A synonym like ``purchase'' would have been
helpful.

\item 
A domain expert can give information about
terms related to particular services. These related terms can be
used as features for quickly 
locating the services. For example, ``name,'' ``address,''
``phone,'' can be some of the relevant terms to the
``customer'' service collection.

\item 
A domain expert can specify certain dependencies between
service collections.
For example, Sales Order is dependent on Customer as it is
likely to use customer details, but  not vice-versa.
Such information can be used to delineate source files specific to
the customer service collection, by removing from them files which
access tables related to the Sales Order service collection.
We have noted in our case-study that these relationships are
strong evidence for discriminating files corresponding to various
service collections. 
\end{enumerate}

\subsection{Useful elements of the application}
\begin{enumerate}
\item 
The directory structure of the application contains useful
information.
Often a collection's implementation is contained in a single directory.
In our case-study we found that every matching collection (out of 9)
was contained in a single directory.

\item 
Database table names associated with a service collection can be of great
help in the matching exercise.
Unlike other code artifacts, tables are not scattered in the
application code (i.e., we can find the database information at single place in most of the systems), and are relatively few in number (119 in \jallin).
Further, database table information (like table names and fields)
are used intact or via macro 
expansions in the source code, and hence are easier to track as
features.

\item 
Much of the \emph{high level} domain-specific terminology (i.e. collection names) used in the application
and the model, was very similar.
This is despite the fact that
the application and model were developed independently.
In our case-study we observed that eight out of nine 
collection names between model and application-UI were similar, while
one pair of names (``material'' and ``item'') were completely
different. Similarly, for the same eight collections a majority of the files corresponding to its implementation had names that resembled the collection names. 

\end{enumerate}

 \section{Heuristics}
\label{sec:heuristics}

In this section  we leverage our observations from the case study to propose some filters. Each of these filters takes the set of source files in the application server side code as input, and outputs a subset of these source files corresponding to each collection in the model. Later, in Section~\ref{ssec:semi-automated-approach}, we present a basic semi-automated approach that uses these filters and matches the model to the code.
We also evaluate the utility of the filters and the heuristic approach by running them on the application under study, and comparing the output with the actual results obtained in our manual study in Section~\ref{section:model-to-code}.

\subsection{Filters}

Each filter takes a set of code features pertaining to the implementation of a collection as a parameter and returns the files in the application's server-side code that have these features.
 Fig.~\ref{fig:collection-table} shows the manually
identified features of the collections that we will make
use of as parameters to the filters described below. The first column of the table lists the collections in the model for which matches were found in the manual study. The second column of the
table shows abbreviated collection names we use in the graphs later in this section.

\begin{figure*}
\small{
\begin{center}
\begin{tabular}{lllll}\hline\\
\textbf{Collection Name ($C$)}× & \textbf{Abbreviated Name}× & \textbf{Tables Accessed ($T_C$)}× & \textbf{Tables
Not-Accessed ($\mathit{TNA}_C$)}×& \textbf{Related Words (${RW}_C$}× \\ \hline\\ 
Customer	× & CUST × & SAL07\_CUSTOMERS	× & DOC01\_SELLING× & name, address, ...× \\
Inbound	Delivery× & IDN ×  & DOC09\_IN\_DELIVERY...× & ---× & delivery, item, ...	× \\
Material (Items)× & ITM ×  & ITM01\_ITEMS, ....	× & DOC01\_SELLING, ....× & product, item, ...	× \\ 
Outbound Delivery× & ODN × & DOC10\_OUT\_...	× & ---× & warehouse, item, ...× \\
ProductionOrder× & PDO × & DOC22\_PRODUCTION...	× & ---× & product, order, ...	× \\
Purchase Order	× & PO ×  & DOC06\_PURCHASE, ...× & ---× & purchase, supplier, ...× \\ 
Sales Order (Sell Order)× & SO ×  & DOC01\_SELLING, ...× & ---× & sales, customer, ...× \\ 
Sales PriceList× & SPL × &  SAL01\_PRICELISTS, ...× & ---× & price, item, ...	× \\ 
Supplier	× & SUPP× & PUR01\_SUPPLIERS	× &DOC06\_PURCHASE, ....× & supplier, item × \\\hline
\end{tabular}
\end{center}
}
\caption{Manually identified features for the collections}
\label{fig:collection-table}
\end{figure*}

\setcounter{paragraph}{0}
\subsubsection{Tables Accessed (TA) Filter}

\begin{figure*}%
\centering
\includegraphics[width=6.0in]{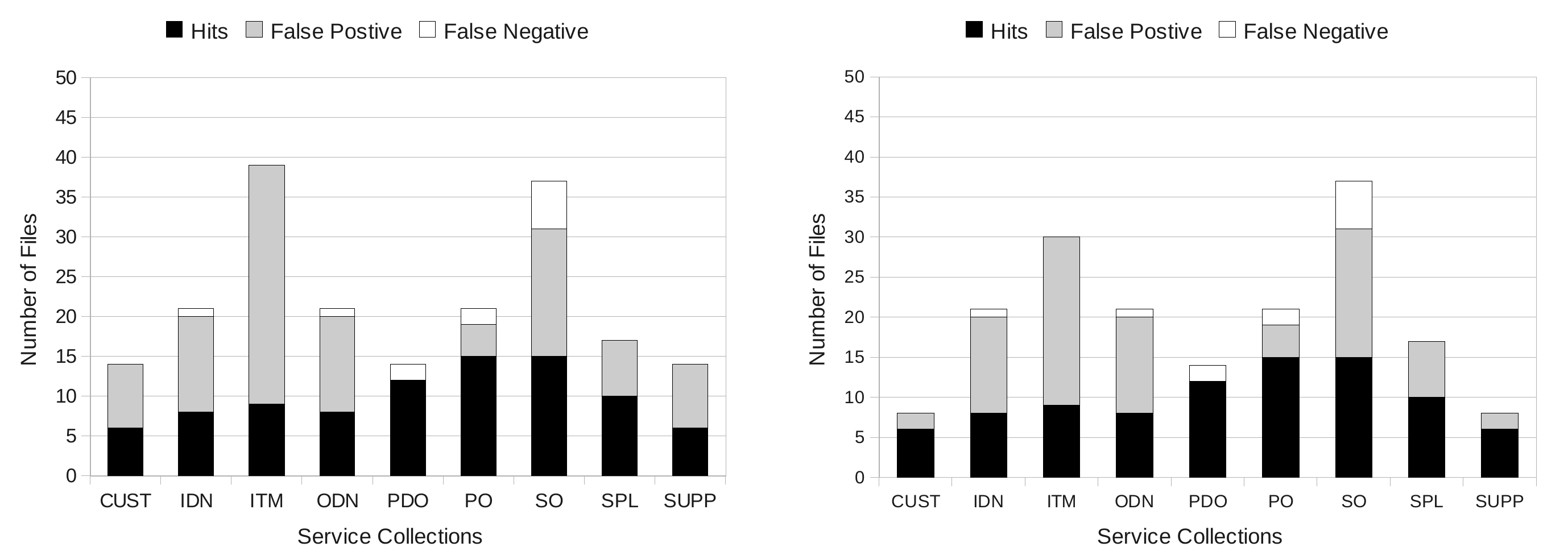}
\caption{Performance of filters TA and ($\mathrm{TNA}\circ\mathrm{TA}$)}
\label{fig:TA-TNA}
\end{figure*}

In our model each collection $C$ correspond to a business entity. Therefore in its implementation, the database tables corresponding to this entity are likely to be accessed. This motivated us to design this filter which takes as parameter a set of core tables $T_C$ (as shown in second column of Fig.~\ref{fig:collection-table}), and returns
the set of all source files that access a table in $T_C$. For example, for the service collection Sales Order, TA filter returns all the source files that access the table ``DOC01\_SELLING''.

 Fig.~\ref{fig:TA-TNA}(a) shows the performance of this filter when given input as shown in the third column of Table~\ref{fig:collection-table}.
The set $F_C$ of source files returned by a filter $F$ is meant to be an approximation of the set $S_C$ of files that actually belong to a collection $C$ (as identified manually in Section~\ref{section:model-to-code}).
For each filter $F$ and collection $C$ we record the number of ``hits''
(i.e., $|F_C \cap S_C|$), the number of
``False Positives'' ($|F_C-S_C|$), and the number of ``False
Negatives'' ($|S_C-F_C|$).
The number of false positives give us an idea of the
``precision'' of the filter, while the number of false negatives
gives us an idea of the ``recall'' of the filter.

We note that we have incomplete recall (namely, for collections IDN, ODN, PDO, PO and SO) along with many false positives (namely, for all collections except PDO).
It should be noted that the precision and recall of a single filter should not be taken as the final precision and recall of the overall approach, since each filter outputs only an approximation of the actual implementation of a collection. We later combine all the filters, with some human intervention, to obtain a good semi-automated matching approach (see Section~\ref{ssec:semi-automated-approach}).

\subsubsection{Tables Not Accessed (TNA) Filter}

This filter is meant to complement the TA filter above. We observed that the TA filter reports many false positives -- source files that access one of the given tables $T_C$ of a collection $C$, but are not part of the collection. An example of this is the CUST collection. Its given table is ``SAL07\_CUSTOMERS.'' However when we run the TA filter with this table as input, it reports (among others) source files pertaining to the Sales Order collection, since the implementations of some services in Sales Order (for example Create Sales Order) access the SAL07\_CUSTOMERS table, to access information related to the customer placing the sales order. Such files, which access both, sales order table and the customer table should correspond to the sales order collection, and not the customer collection. We found a similar relationship for two other collections SUPP and ITM (with the implementation of ITM dependent on SUPP, but not vice-versa).

The filter TNA is motivated by such scenarios. It takes as parameter the set of tables $\mathit{TNA}_C$ \emph{not} accessed by the collection $C$, and outputs all source files that do not access any of the tables in $\mathit{TNA}_C$. 
Thus, in the case of the CUST service collection above, we include ``DOC01\_SELLING'' in the input $\mathit{TNA}_{CUST}$ to the filter.

Fig~\ref{fig:TA-TNA}(b) shows the results of the TNA filter applied to the results of the TA filter, on the application studied.
As indicated in Fig.~\ref{fig:collection-table} we had ``tables not accessed'' information for the collections CUST, ITM, and SUPP.
Note the reduction in false positives we get for these three collections by applying this filter.

The TNA filter is just a representative of the filters which can be designed based on the relationships among entities present in the domain. As we see from the results of this filter even a very basic relationship can be exploited to reduce false positives (i.e., to increase precision). Therefore it will be useful if domain experts could provide more information about the relationships among the business entities.

The next three filters, described below, help mainly with the process of
``expanding'' the files around the seed files to obtain the complete set of
source files corresponding to a given collection. These filters assign a
score to each remaining non-seed file, and output source files that have a
score above a given threshold. The files in the output have high
likelihood of being pertinent to the collection.

\subsubsection{Filename (FN) Filter}

\begin{figure*}%
\centering
\includegraphics[width=6.0in]{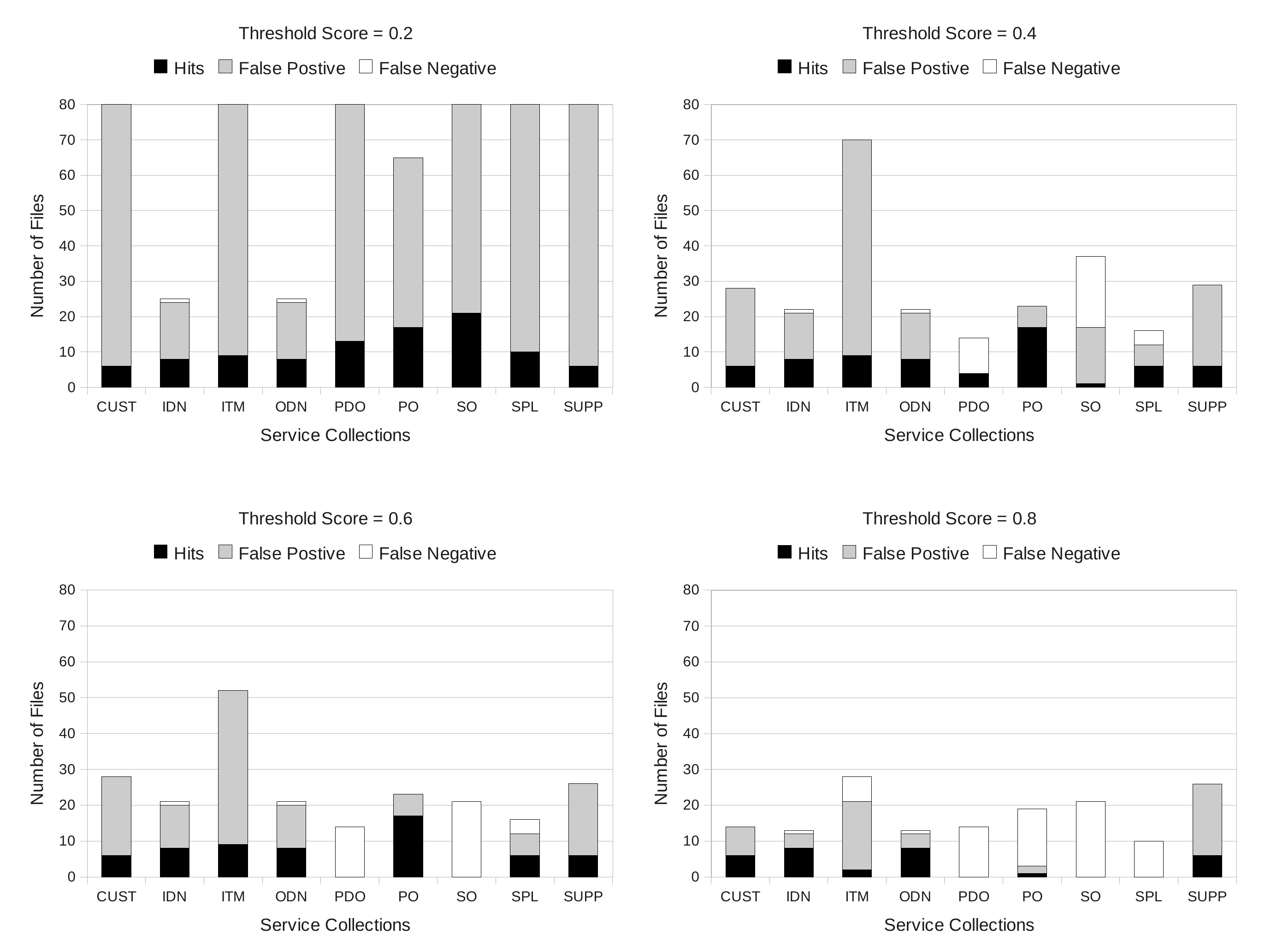}
\caption{FN Filter with matching threshold values of 0.2, 0.4,
  0.6, and 0.8. Maximum number of files shown is 80.}
\label{fig:FN}
\end{figure*}

This filter tries to exploit the fact that the names of source files
corresponding to a collection $C$ are often closely related to the
canonical name used in the implementation to refer to the main entity
operated upon by the collection.  For example by examining the tables in
$T_{ITM}$ associated with the Material collection (see third row, third
column in Figure~\ref{fig:collection-table}), we inferred that ``material''
in the model is referred to canonically as ``item'' in the implementation. Similarly,
``sell order'' is used in the implementation to refer to ``Sales Order'' in
the model. 

For each collection $C$, the FN filter works by first concatenating the
words in the canonical implementation-name of this collection to obtain a
string $s$, and then gives a score to each file, which is the length of the
longest substring of $s$ that occurs in the name of the file divided by the
length of $s$ itself. It then outputs the files whose score is above $v$,
where $v$ is a threshold parameter between 0 and 1. The canonical
implementation-name is given as a parameter to the filter, as shown within
parenthesis (wherever it differs from the name of the collection in the
model) in Column~1 in Figure~\ref{fig:collection-table}.

Fig.~\ref{fig:FN} shows the results of the FN filter with matching threshold values of 0.2, 0.4, 0.6, and 0.8 respectively. We note that with a threshold value of 0.4 we get good precision and recall.

From the results of this filter we infer that the programmers do not use arbitrary names for various elements of application e.g., function names, file names, variable names etc.. They use abbreviated forms of the domain terms as part of the names of various program entities. But we need to synchronize with the programmers' terminology which we can do to a good extent by using database information as we do in case of Material collection.

\subsubsection{Table Fields (TF) Filter}

\begin{figure*}%
\centering
\includegraphics[width=6.0in]{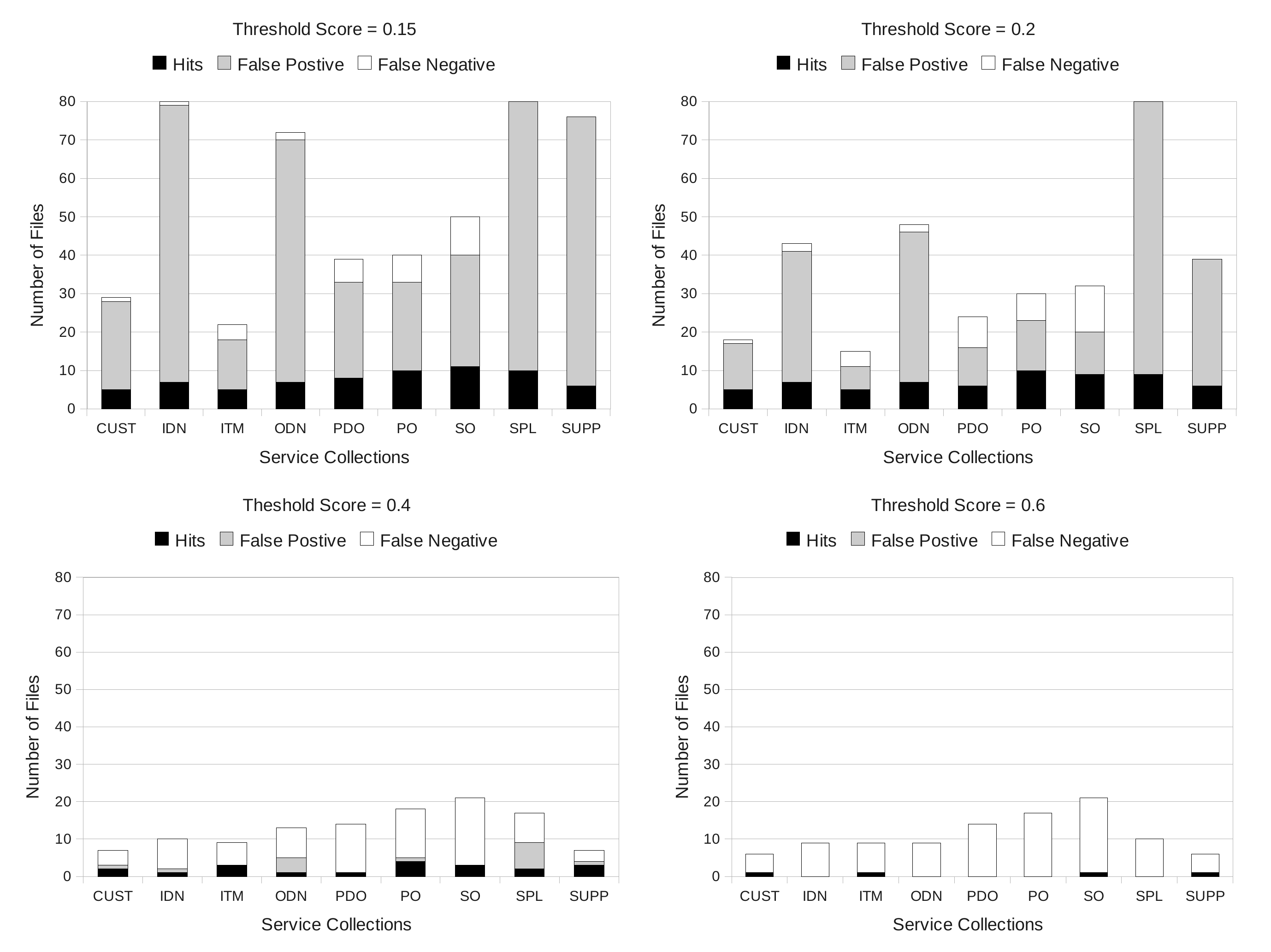}
\caption{TF Filter with threshold percentages of 15\%, 20\%,
  40\%, and 60\% respectively. Maximum number of files shown is 80.}
\label{fig:TF}
\end{figure*}

The motivation for this filter is if a large percentage of the fields of a
table $t \in T_C$ pertinent to a collection $C$ are accessed in a file, but
the table itself is not accessed in the file, then the file is pertinent to
the collection. This happens, e.g., when a method in the file
receives a row of data as a parameter from another file that accesses the
table, and the method processes or prepares this row (by referring to the
fields in the row). 

This filter first gives a score to each file, which is the percentage of
fields accessed of all the tables in $T_C$. The filter outputs 
source files whose score is more than a threshold parameter $v$.

Fig.~\ref{fig:TF} shows the performance of this filter on the application under study.

We observe from the results of this filter that for some of the collections at a greater percentage (i.e., 20\% or more) of total number of fields, we get fewer files. Whereas for a few of the collections a small percentage gives a small number of files in the output. It is because of the density of fields i.e., if the total number of fields is very large then only a small percentage of the fields will be accessed by a file or the functionality, whereas if total number of fields are small then most of the fields are expected to be accessed. Overall after tuning the threshold for each collection to output fewer  files, we find the resulting files give moderate recall and precision. 


\subsubsection{Related Words (RW) Filter}

\begin{figure*}%
\centering
\includegraphics[width=6.0in]{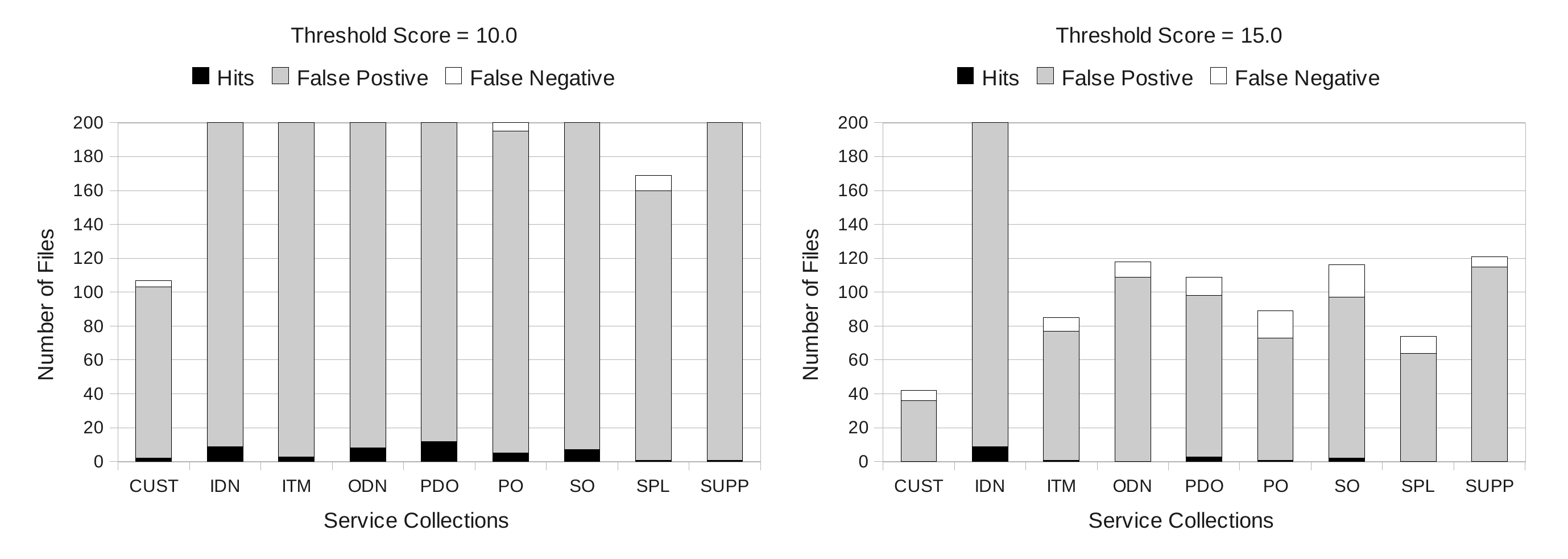}
\caption{RW Filter with threshold 10.0 and 15.0. Maximum number of files shown is 200.}
\label{fig:RW}
\end{figure*}

The motivation for this filter is that the occurrence of words related to a collection $C$'s name in a source file is evidence that the file belongs to the implementation of collection $C$.

This filter takes a set of related words corresponding to each collection as input, as shown in the last column of the Fig.~\ref{fig:collection-table}. We populate the related words for the collections manually, using domain knowledge.
For example, for the Customer collection, we may use the set of related words like ``customer,'' ``name,'' ``address,'' ``city,'' and ``country.'' This filter outputs a set of files containing some percentage of approximately matching words.

This filter also assigns a score to each file $f$ corresponding to a given collection $C$, and then outputs the files having score equal to or above a given threshold value $v$. The score corresponds to the average number of approximate accesses to each word in the set ${RW}_C$. For calculating this value, we first take each file $f$ as a bag of words and matches each word $w_f$ in the file $f$ with every word in ${RW}_C$. If the division of, length of longest substring between $w_f$ and a word $w_R$ in ${RW}_C$, and the length of word $w_R$, is equal to or greater than a given threshold value $u$, then we increment the count of the number of approximate accesses $A_n$ to the words in ${RW}_C$ by the file $f$. Finally score is computed by dividing the count $A_n$ with $|{RW}_C|$. In our experiments we set 0.6 as the value of $u$. 

The results of this filter are shown in Fig.~\ref{fig:RW}. 

We observed that both, the precision and recall of the filter were poor. We see a reason for this, that our set of related keywords was small (due to lack of domain expertise), and also that the set of related keywords was not consistent with the terminology used in the programs. 



We note an important issue regarding the filters defined above. They all assume a file-level granularity -- that is, each source file belongs to at most one service collection. Techniques such as concept assignment along with slicing~\cite{HarmanMark2002} may be required to address the scenario wherein service implementations are at finer granularities.

\subsection{A Semi-Automated Approach} 
\label{ssec:semi-automated-approach}
We now suggest a semi-automated algorithm, using the filters defined
above, to partially automate the procedure of matching the model of
service descriptions to the source code. The algorithm is shown in Fig.~\ref{fig:sa-algo}.

\begin{figure}
\rule{3.5in}{0.5pt}
\begin{enumerate}
  \item User creates a table of features, as in Fig.~\ref{fig:collection-table}.
  \item For each collection $C$, apply the TA filter, followed by the TNA filter, to obtain a set of 
	candidate seed files $C_C$.
  \item For each file $f$ in $C_C$, add $f$ to an initial ``seed'' source files set $M_C$,
    \begin{enumerate}
    \item if $f$'s score due to FN filter is in the top $X_1\%$ of scores among all files in $C_C$, or
    \item if $f$'s score due to TF filter is in the top $X_1\%$ among all files in $C_C$, or
    \item if $f$'s score due to each of the filters FN and TF is in the top $Y_1\%$ of scores among all files in $C_C$.
    \end{enumerate}
  \item Create a temporary set $C_T$ and add all files in $M_C$ to it. (We use $C_T$ 
  to store candidate files for expansion.)
  \item Do (perform expansion)
    \begin{enumerate}
     \item Apply the FN filter and tune the threshold value automatically so that at most $X_3$ number of files are returned by the filter.
	  Add a file among these to $C_T$, if it is an immediate neighbour in the call graph of a file already in $C_T$.
     \item Apply the TF filter similarly.
    \end{enumerate}
	While no new files are added.
  \item Create a candidate expansion set, $C_E = C_T - M_C$.
  \item For each file $f$ in $C_E$, add $f$ to $M_C$,
    \begin{enumerate}
    \item if $f$'s score due to FN filter is in the top $X_2\%$ of scores among all files in $C_E$.
    \item else if $f$'s score due to TF filter is in the top $X_2\%$ among all files in $C_E$.
    \item if $f$'s score due to each of the filters FN and TF is in the top $Y_2\%$ of scores among all files in $C_E$.
    \end{enumerate}
  \item{\label{step:manual}} Manually analyze files in $M_C$ and remove irrelevant files from it.
  \item Expand $M_C$ manually, by adding to it other files which are closely related to $M_C$ in the call graph and are highly relevant to $C$.
  \item If a file $f$ is called by or calls files in $M_C$ only, and is not contained in the set $M_D$ of any other collection $D$, then add $f$ to $M_C$.
  \item Return $M_C$ as the output set implementing the given collection $C$.
\end{enumerate}
\rule{3.5in}{0.5pt}
\caption{A semi-automated algorithm to identify the implementation of a collection $C$, 
using TA, TNA, FN and TF filters.}
\label{fig:sa-algo}
\end{figure}

\begin{figure}
\centering
\includegraphics[width=3.5in]{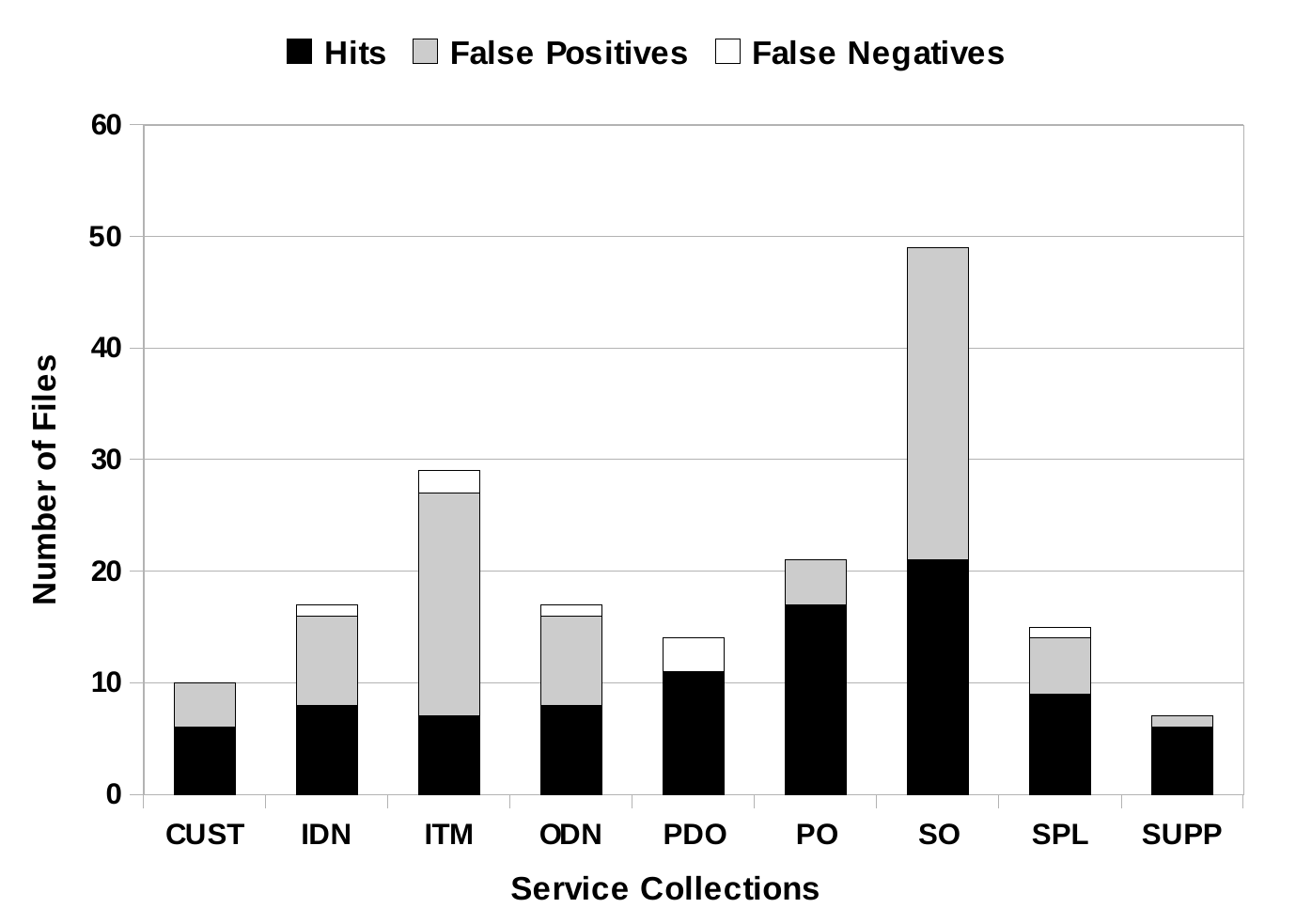}
\caption{Performance of semi-automatic approach}
\label{fig:semi-automated}
\end{figure}

Steps 2 to 7 of the algorithm are automated, whereas steps 1, and 8 to 10 are manual. The results we report in this section as shown in Fig.~\ref{fig:semi-automated} were obtained by omitting steps 8 to 10. We used, $20\%$, $80\%$ and $10\%$ as the values of $X_1$, $Y_1$ and $X_2$ respectively, and did not use the Step~7(c). We set $X_3$ to $70$ to tune the FN and TF filters in Step 5(a) and Step 5(b). We roughly guessed this number (70) for tuning based on (a)~the possible number of collections in the application, which we approximated based on database tables (119 tables in JAllInOne), and (b) the application size (1089 files in JAllInOne). 

We show the overall precision and recall of the algorithm in Fig.~\ref{fig:semi-automated}, and the actual set of files identified by it for each collection in the second column of Fig.~\ref{fig:allfiles}.

The precision of the algorithm ranged from 100\% (for the collection PDO)
to 26\% (for the collection ITM). While investigating this issue we found
that most of the false positives
were passing through the FN filter. Collections ITM and SO are the most affected in this way. 
In the case of ITM, the collection name in the implementation (i.e., ``items'')
is small; therefore, several irrelevant files also get approximately the
same score as relevant files. We notice a large number of false positives in case of SO collection also, as ``order'' term of ``sell-order'' is larger than ``sell'', and matches with several other file names containing ``order'' term.
Note that the recall of the algorithm is 89\%, 78\%. 89\%, 79\%, 90\%, and
100\%, for IDN, ITM, ODN, PDO, SPL, and all other collections,
respectively. 

For collection ITM, the algorithm misses 2 files. It is due to our selection of
values for the parameters used in Step 3 and Step 7. These missed files
(false negatives) have names that match poorly with the name of the
collection, and access a table that has fewer  fields than most
other tables. Therefore these two files do not pass the test in Step 3. Two
false negatives occurred for collection ``Production Order'' for similar
reasons. These kinds of files can be found by designing a more
sophisticated TF filter that assigns separate scores to a file for each
related table and then combines these scores in a meaningful way to assign
a final score to the file. Some of other false negatives (e.g, one for
``Indelivery Notes'', one for ``Outdelivery Notes'', one for ``Production
Order'', etc.) come about because we did not use Step~10 while reporting the
results.

To summarize, in this section we described and evaluated a preliminary semi-automated algorithm for matching collections to their implementations. With carefully chosen tuning parameters we were able to achieve satisfactory performance with respect to recall. More work is required to investigate techniques to improve the precision of the approach.


%

 \section{Related Work}
\label{sec:related-work}

Mining services from monolithic applications is a well-researched area, with a number of approaches published in the literature. These approaches can be classified broadly into two categories, that we term as bottom-up service mining and top-down service mining. In bottom-up service mining, the focus is on extracting high level components from source code and wrapping them as services \textit{without} a prior model of the required services. In the top-down service mining, which is the approach followed in our work, the focus is on the business domain and on identifying functionality in the monolithic code matching abstract service descriptions in the model created by the business domain experts.

The role of user recognizable components (i.e. concepts or services) is described in the work of Rajlich and Wilde~\cite{RajlichVaclav2002}. They discuss how the concepts (or services) help users in understanding the architecture of the software. They also present the benefits of concept oriented program comprehension and locating concepts instead of arbitrary high level components. Concepts promote reuse also, since arbitrary clusters of software modules do not have significant meaning for users and therefore are difficult to reuse. A case study presented by Haiduc and Marcus~\cite{haiduc2008use} show that the domain terms are generally extensively used in the applications. In their study they use the graph theory domain. They found in their study that 42\% domain terms were present in source code, out of which 23\% were present in the comments alone. But as we observed in our study that JAllInOne (application under study) has only one or two lines of meaningful comments (may be even lesser in legacy systems) and there were very few matching domain terms from the domain model contained in these. It shows that the richness of comments and domain terms in application may be dependent on domain or size. It may also have to do with the fact that we used a real domain model that was developed independently of the application, and did not contain too many low-level domain terms, whereas the domain model of Rajlich et al. was probably created by the authors.

A number of approaches have been discussed in literature as bottom-up approaches on identifying potential services from source code. These include techniques based on software clustering \cite{zhang2005service,Zhang04}, graph analysis \cite{Li2006}, software metrics \cite{victor91}, formal concept analysis \cite{Antoniol2001} and analysis of design documents \cite{Kannan2008}. One drawback however is that, since bottom-up approaches do not start from a model of the required services, the granularity and functionality of the services identified depends on the underlying technology used, and hence may not match the granularity and functionality as required by the architect. In our approach, while we have primarily followed a top-down approach, we have used aspects of clustering and slicing to improve the precision. 

Grosso et al.~\cite{DelGrosso2007} infer a service for each database query contained in the source code. Their services may or may not be user recognizable. In fact as we observed in our case study that a service may issue more than one database query and that every query need not necessarily correspond to a domain service. 

Clustering and data mining techniques have also been used by a series of papers including ~\cite{Zhang04, zhang2005service, Carey2007} to modularize system or to identify the high level components. The pioneer work of Wiggerts~\cite{wigg97} gives the overview of most of clustering approaches in software context. They also discuss the components of a software which can be considered as the entities to be clustered and the types of similarity measures among them. All the bottom-up approaches based on software clustering suffer with the problems associated with bottom-up approaches like service-naming and granularity of services.

Another approach in the bottom-up category, abstracting the classes as graph nodes and call-edges as the edges among nodes is used by Li and Tahvildari~\cite{Li2006}. In this technique they present each entry point as a top-level service and later by doing some graph transformations other components also as low-level services. Later they show these services to users to identify useful ones and assign meaningful names. As the components identified are independent of user understanding of services (domain concepts), therefore it may be difficult for the users to assign meaningful names to the components. Also, their technique doesn't guarantee that the components necessarily expose the services of adequate granularity. Whereas our technique is uses the user directed concepts (i.e., business entities or concepts) to mine service implementations, and therefore all the mined implementations correspond to user recognizable services. 

Briefly, in the bottom up approaches the goal is generally to modularize the system as high level components, which are not necessarily user recognizable. On the other hand our goal is to identify user-recognizable high level components. The problem with these approaches is also, that once they identify some high level components without prior model then users need to go through all these components to assign meaningful names. Another problem is the granularity of services, most of the time the granularity of services identified by bottom-up approaches do not match with the user recognizable granularity. They end up with either finer or coarser granularity. 


The second category, i.e., top-down service mining involves matching natural language descriptions in a model or query with source code artifacts. This comes under the purview of the areas of concept assignment and feature location, which predominantly use information retrieval (IR) techniques for locating source code matching a given description expressed in domain vocabulary. The use of Information Retrieval for service identification started with the work of Biggerstaff et al.~\cite{bigg93}. In their approach, based on the features corresponding to domain services (they use the term ``concept") they mark manually the contiguous segments of code and present to users as the implementing components. In the same year Lanubile et al.~\cite{lanu93} proposed the use of slicing for finding the executable components corresponding to various services. In their approach the user needs to mark the statements corresponding to services required to be identified and by extracting slices corresponding to these marked statements they find the components. Harman et al.~\cite{HarmanMark2002} later unified both and gave an approach to extract executable slices corresponding to the domain services. Our goal is different here in that instead of identifying executable components our goal is to find the service implementations having non-overlapping boundaries with other service implementations, possibly by invoking services from other implementations for its execution. 

The work of Sindhgatta and Ponnalagu~\cite{sindhgatta2008locating} is closely related to ours. They propose an approach to locate components that realize the services in  existing systems. Their approach mainly involves two steps; firstly they extract links between service descriptions and source code implementations. For this they use the terms available in the model, and input/output to the various services. They use Information Retrieval methods (specifically use the tool Lucene) to match the model terms with the program modules. In the second step they use structural dependencies and some metrics to rank the initial links, and ask the user to remove some of these links for precision. The overall approach can be seen as first finding the over-approximation of the implementation, and then contraction with human support for precision. In contrast, in our methodology we first find the seeds and then expand to find the full implementation. There are two other differences between our work and theirs. The first is, we have an extensive manual case study, the results of which we use to derive a heuristic. Secondly, we use a real domain model, developed independently of the code. One observation we made as a result of this, which is in contrast to what they observe, is that there is not much overlap in domain terms between the model and the code. Our observation was that finding the database tables that are pertinent to a collection or not pertinent to it (see filters TA and TNA) was a better way of matching the model to the code.  

Dynamic feature location is the technique proposed in the work of Wilde and Scully~\cite{wilde95}. Dynamic feature location basically takes dynamic traces corresponding to the test cases designed for various features. Then, by taking differences of these traces figures out the computational units (code components) corresponding to a particular feature (or service). This work is later followed by Eisenbarth et al.~\cite{Koshke03}, in which they use a combination of dynamic analysis, static analysis and concept lattice for a locating features in source code. They use the term \textit{feature} for what we call services. First, they create scenarios and then generate traces corresponding to these scenarios. By taking difference of the execution traces and using static analysis find the computational units (i.e. components realizing the services). Then they create a concept lattice (using Formal Concept Analysis - FCA~\cite{wille82}) by considering scenarios as attributes and computational units as objects. Based on this concept lattice they associate a maximal set of objects with every set of attributes in the lattice. Later, find the computational units (code modules) corresponding to various sets of attributes and associate them with the services. For this approach to be useful in our setting first of all we need to find the precise set of classes (or functions) corresponding to the various scenarios, but as we showed in our manual methodology (Fig.~\ref{fig:so-ex-trace}) that the execution traces generally do not correspond to the specific features or services. They contain other classes corresponding to supporting classes also which make it difficult to associate the classes (or files) to any particular service.

There are several other papers in this area which predominantly use Information Retrieval techniques to address the problem of software maintenance and a few of them aim to identify implementations corresponding to the service descriptions. One of this kind of approach is given by Marcus et al.~\cite{Marcus2004} in which they use LSI (Latent Semantic Indexing) to find similarities between model and the source code modules. It is mainly based on the terms available in the model and does not use any other structural information related to programs, like call graph, program flow etc.. Several other approaches are given by researchers in the literature~\cite{Kannan2008, Carey2007, Gay2009, HarmanMark2002, Marcus2004, RajlichVaclav2002, Robillard2007b} to locate service implementations mainly using related terms along with some program flow information. But as we show in  Section~\ref{sec:heuristics} that if the model or query is not rich enough in related terms and if it is not in sync with the developers terminology then the results may be not be good. As we understand that enriching the model or query in related terms and synchronizing these with programmers terminology is quite difficult for the domain experts writing the model. Whereas our methodology suggests use of heuristics which are based on database tables and terminology extracted from database information only to locate the service implementations instead of relying only on the related terms given in model or query.

%

\begin{figure*}%
\scriptsize{
\begin{center}
\begin{tabular}{lll}\hline\\
\textbf{Abbrev.}× & \textbf{Manually Identified}× & \textbf{Heuristic-Algorithm}× \\ 
\textbf{Name $C$}× & \textbf{Implementation ($S_C$)}× & \textbf{Output ($M_C$)}× \\
× & 	× & 	× \\\hline
× & 	× & 	× \\
CUST ×  & DeleteCustomersAction, UpdateCustomerAction, × & 	UpdateTaxableIncomesBean, LoadCustomerAction, DeleteCustomersAction, × \\
× & 	ValidateCustomerCodeAction, InsertCustomerAction, × & 	InsertCustomerAction, LoadCustomersAction, 	UpdateCustomerAction,× \\
× & 	LoadCustomersAction, LoadCustomerAction× & 	ValidateCustomerCodeAction, LoadOutDeliveryNoteAction, × \\
× & 	× & 	CreateSaleDocFromEstimateAction, LoadOutDeliveryNotesAction, × \\
× & 	× & 	× \\
IDN ×  & InsertInDeliveryNoteRowAction, LoadInDeliveryNotesAction,× & UpdateInDeliveryNoteAction, DeleteInDeliveryNoteRowsAction, × \\
× & 	LoadInDeliveryNoteRowsAction, InsertInDeliveryNoteAction,× & 	InsertOutDeliveryNoteAction, LoadInDeliveryNotesAction, × \\
× & 	 UpdateInDeliveryNoteAction, DeleteInDeliveryNoteRowsAction,× &  InsertInDeliveryNoteAction, LoadOutDeliveryNotesAction, × \\
× & 	 InsertInSerialNumbersBean, LoadInDeliveryNoteAction,× &  LoadOutDeliveryNotesForSaleDocAction, InsertOutDeliveryNoteRowAction, × \\
× & 	 UpdateInDeliveryNoteRowsAction× & LoadInDeliveryNotesForPurchaseDocAction, LoadInDeliveryNoteRowsAction, × \\
× & 	× & LoadInDeliveryNoteAction, InsertInDeliveryNoteRowAction, × \\
× & 	× & 	LoadOutDeliveryNoteAction, UpdateInDeliveryNoteRowsAction, × \\
× & 	× & 	UpdateOutDeliveryNoteAction, LoadOutDeliveryNoteRowsAction, × \\
× & 	× & 	× \\
ITM ×  & InsertItemAction, LoadItemAction, × & InsertItemAction, LoadSupplierItemsAction, UpdateSaleDocRowAction,× \\ 
× & 	InsertItemAttachedDocsAction, DeleteItemsAction, × & 	LoadMovementsAction, LoadPriceItemsAction, ValidateCustomerCodeAction, × \\
× & 	UpdateItemAction, LoadItemsAction, × & 	LoadPurchaseDocAndDelivNoteRowsAction, ValidateVatCodeAction, LoadItemAction, × \\
× & 	LoadItemAttachedDocsAction, ValidateItemCodeAction, × & UpdateItemAction, ValidatePriceItemCodeAction, LoadItemImplosionAction, × \\
× & 	DeleteItemAttachedDocsAction× & 	ValidateItemCodeAction, ValidateSupplierItemCodeAction, LoadItemsAction, × \\
× & 	× & 	LoadSupplierPriceItemsAction, LoadBillOfMaterialBean, ImportAllItemsAction, × \\
× & 	× & 	LoadItemAvailabilitiesAction, LoadOrderedItemsAction, DeleteItemsAction, × \\
× & 	× & 	ValidateSupplierPriceItemCodeAction, CreateInvoiceFromScheduledActivityAction, × \\
× & 	× & 	LoadScheduledItemsAction, ImportAllItemsToSupplierAction, × \\
× & 	× & 	LoadCallOutItemsAction, LoadItemAttachedDocsAction × \\
× & 	× & 	× \\ 
ODN × & InsertOutSerialNumbersBean, UpdateOutDeliveryNoteAction, × & UpdateInDeliveryNoteAction, DeleteInDeliveryNoteRowsAction, × \\
× & 	InsertOutDeliveryNoteRowAction, LoadOutDeliveryNotesAction, × & InsertOutDeliveryNoteAction, LoadInDeliveryNotesAction, LoadInDeliveryNoteAction, × \\
× & 	DeleteOutDeliveryNoteRowsAction, LoadOutDeliveryNoteAction, × & InsertInDeliveryNoteAction, LoadOutDeliveryNotesAction, InsertInDeliveryNoteRowAction, × \\
× & 	LoadOutDeliveryNoteRowsAction, UpdateOutDeliveryNoteRowsAction, × & LoadOutDeliveryNotesForSaleDocAction, InsertOutDeliveryNoteRowAction, × \\
× & 	InsertOutDeliveryNoteAction × & 	LoadInDeliveryNotesForPurchaseDocAction, LoadInDeliveryNoteRowsAction, × \\
× & 	× & 	LoadOutDeliveryNoteAction, UpdateInDeliveryNoteRowsAction, × \\
× & 	× & 	UpdateOutDeliveryNoteAction, LoadOutDeliveryNoteRowsAction× \\
× & 	× & 	× \\
PDO × & LoadProdOrderProductsAction, UpdateProdOrderAction, × & DeleteProdOrderProductsAction, UpdateProdOrderAction, × \\
× & 	CloseProdOrderAction, CheckComponentsAvailabilityAction, × & 	CloseProdOrderAction, InsertProdOrderAction, × \\
× & DeleteProdOrderProductsAction, InsertProdOrderAction,× & LoadProdOrderComponentsAction, LoadProdOrderAction, × \\
× & 	LoadProdOrderAction, LoadProdOrderComponentsAction, × & UpdateProdOrderProductsAction, ConfirmProdOrderAction, × \\
× & 	InsertProdOrderProductsAction, DeleteProdOrdersAction, × & LoadProdOrderProductsAction, InsertProdOrderProductsAction, × \\
× & ConfirmProdOrderAction, CheckComponentsAvailabilityBean,× & LoadProdOrdersAction × \\
× & 	UpdateProdOrderProductsAction, LoadProdOrdersAction× & 	× \\
× & 	× & 	× \\
PO ×  & UpdatePurchaseDocAction, PurchaseDocTotalsBean, × & 	LoadPurchaseDocAction, ConfirmPurchaseOrderAction, × \\
× & 	ConfirmPurchaseOrderAction, PurchaseDocTaxableIncomesBean, × & 	ValidatePurchaseDocNumberAction, UpdatePurchaseDocRowAction, × \\
× & 	DeletePurchaseDocsAction, LoadPurchaseDocRowAction, × & DeletePurchaseDocsAction, ClosePurchaseDocAction, UpdatePurchaseDocAction,× \\
× & 	InsertPurchaseDocAction, ValidatePurchaseDocNumberAction, × & 	InsertPurchaseDocAction, LoadPurchaseDocRowAction, × \\
× & 	ClosePurchaseDocAction, LoadPurchaseDocBean, × & 	LoadPurchaseDocRowsAction, LoadPurchaseDocAndDelivNoteRowsAction, × \\
× & 	LoadPurchaseDocRowsAction, PurchaseDocTotalsAction, × & LoadInDeliveryNotesForPurchaseDocAction, PurchaseDocTaxableIncomesBean, × \\
× & 	LoadPurchaseDocsAction, InsertPurchaseDocRowAction, × & CreateInvoiceFromPurchaseDocAction, LoadPurchaseDocBean, 	× \\
× & 	DeletePurchaseDocRowsAction, UpdatePurchaseDocRowAction, × & PurchaseDocTotalsAction, DeletePurchaseDocRowsAction, × \\
× & 	LoadPurchaseDocAction× & 	PurchaseDocTotalsBean, LoadPurchaseDocsAction, × \\
× & 	× & 	UpdateInQtysPurchaseOrderBean, InsertPurchaseDocRowAction, × \\
× & 	× & 	× \\
× & 	× & 	× \\
SO ×  & ConfirmSaleDocAction, ValidateSaleDocNumberAction, × & DeleteSaleDocRowsAction, LoadSaleDocRowsAction, InsertSaleDocRowBean, × \\ 
× & 	LoadSaleDocRowsBean, LoadSaleDocRowsAction, × & 	InsertSaleDocBean, InsertSaleDocChargesAction, LoadSaleDocActivitiesBean, × \\
× & 	LoadSaleDocRowAction, InsertSaleSerialNumbersBean, × & 	ValidateSaleDocNumberAction, SaleDocTaxableIncomesBean, LoadSaleDocRowBean, × \\
× & 	DeleteSaleDocsAction, DeleteSaleDocRowsAction, × & 	SaleItemTotalDiscountBean, InsertSaleDocActivityBean, CloseSaleDocAction, × \\
× & 	CloseSaleDocAction, UpdateSaleDocRowAction, × & 	UpdateSaleDocTotalActivityBean, InsertSaleDocRowAction, LoadSaleDocActivitiesAction, × \\
× & 	LoadSaleDocAndDelivNoteRowsAction, LoadSaleDocsAction,× & 	ConfirmSaleDocAction, SaleDocTotalsAction, CreateSaleDocFromEstimateAction, × \\
× & 	 InsertSaleDocAction, CreateSaleDocFromEstimateAction, × & 	LoadSaleDocDiscountsBean, InsertSaleSerialNumbersBean, LoadSaleDocDiscountsAction, × \\
× & 	LoadSaleDocRowBean, LoadSaleDocBean, InsertSaleDocRowBean,× & 	UpdateOutQtysPurchaseOrderBean, LoadOutDeliveryNotesForSaleDocAction, × \\
× & 	InsertSaleDocRowAction, LoadSaleDocAction, × & 	InsertSaleDocDiscountsAction, LoadSaleDocAndDelivNoteRowsAction, UpdateSaleDocAction,× \\
× & 	UpdateSaleDocAction, InsertSaleDocBean × & 	UpdateOutQtysSaleDocBean, SaleItemTotalDiscountAction, ExportRetailSaleOnFileBean, × \\
× & 	× & InsertSaleDocDiscountBean, InsertSaleDocRowDiscountsAction, 	LoadSaleDocRowAction, × \\
× & 	× & 	UpdateSaleDocTotalDiscountBean, LoadSaleDocBean, LoadSaleDocsAction, × \\
× & 	× & UpdateSaleDocTotalChargeBean, InsertSaleDocRowDiscountBean, UpdateSaleDocRowAction, × \\
× & 	× & UpdateSaleItemTotalDiscountBean, DeleteSaleDocsAction,  LoadSaleDocChargesBean,× \\
× & 	× & 	InsertSaleDocChargeBean, LoadSaleDocChargesAction, LoadSaleDocRowsBean, × \\
× & 	× & 	CreateInvoiceFromSaleDocAction, InsertSaleDocActivitiesAction, × \\
× & 	× & 	LoadSaleDocAction, InsertSaleDocAction, LoadSaleDocRowDiscountsAction× \\
× & 	× & 	× \\ 
SPL × & ValidatePricelistCodeAction, ChangePricelistAction, × & ValidatePriceItemCodeAction, ValidatePricelistCodeAction, × \\
× & 	UpdatePricesAction, LoadPricelistAction, × & 	ChangePricelistAction, LoadPriceItemsAction, InsertPricelistsAction, × \\
× & 	InsertPricesAction, UpdatePricelistsAction, × & ValidateSaleDocNumberAction, InsertPricesAction, UpdatePricesAction, × \\
× & 	InsertPricelistsAction, LoadPricesAction, × & 	LoadSaleDocBean, LoadSaleDocsAction, 	UpdatePricelistsAction,× \\
× & 	DeletePricesAction, DeletePricelistAction× & 	DeletePricelistAction, LoadPricelistAction, LoadPricesAction× \\
× & 	× & 	× \\
SUPP× & InsertSupplierAction, ValidateSupplierCodeAction, × & 	UpdateSupplierAction, ValidateSupplierCodeAction, × \\
× & 	LoadSuppliersAction, LoadSupplierAction, × & 	LoadSuppliersAction, InsertSupplierAction, LoadSupplierAction,× \\
× & 	DeleteSuppliersAction, UpdateSupplierAction× & 	LoadSupplierPricesAction, DeleteSuppliersAction × \\ 
× & 	× & 	× \\
× & 	× & 	× \\\hline
\end{tabular}
\end{center}
}
\vspace*{2pt}
\caption{Manually identified implementation ($S_C$), and Heuristic algorithm output ($M_C$), for each of the nine collections}
\label{fig:allfiles}
\end{figure*}

 \section{Future Work}
\label{sec:future-work}

In the future we would like to undertake more case studies to evaluate the
performance of our methodology and heuristics on more varied applications
and domain models. In particular, we would like to use larger, more
complete domain models (perhaps from domains other than ERP), as well as
larger applications. We will continue our efforts to locate legacy
applications, in order to tune our approach to the idioms they normally
exhibit. We will try to improve the rigor of our case study and reduce
subjectivity by having independent experts review the results of our
automated heuristics.

Parts of the methodology like seed-finding and expansion need to be better
understood, and automated further.  We intend to explore program analysis
techniques to recover more and richer features from the code, that will
assist the matching.  We intend to explore information-retrieval techniques
to both reduce the burden on humans to provide inputs (e.g., related tables
and related words for each collection), and to resolve ambiguities in
finding the seeds of and fixing the boundaries of collection- and
service-implementations.  We need to investigate intelligent ways to
interactively accept human insight during the semi-automated matching
process.  Finally, we would like to study alternative techniques reported
in the literature, and to see if we can incorporate those techniques as
well as our own in a combined system (e.g., in a probabilistic belief
system).

 \section{Conclusion}
\label{sec:conclusions}

In this report we presented a novel three-way matching and validation
methodology to identify implementations of desired collections and services
in an existing application.  We pin-pointed the issues which require
subjective decisions during the matching, as well as code-features
that developers ought to provide as input in order to make the methodology
effective. We made a set of observations from our study (e.g., we found that TA and TNA filters are the most effective for matching, in contrast with the domain terms which do not match much), and then designed
a semi-automated approach based on these observations.

\bibliographystyle{IEEEtran}

\bibliography{library}

\begin{thebibliography}{10}
\providecommand{\url}[1]{#1}
\csname url@samestyle\endcsname
\providecommand{\newblock}{\relax}
\providecommand{\bibinfo}[2]{#2}
\providecommand{\BIBentrySTDinterwordspacing}{\spaceskip=0pt\relax}
\providecommand{\BIBentryALTinterwordstretchfactor}{4}
\providecommand{\BIBentryALTinterwordspacing}{\spaceskip=\fontdimen2\font plus
\BIBentryALTinterwordstretchfactor\fontdimen3\font minus
  \fontdimen4\font\relax}
\providecommand{\BIBforeignlanguage}[2]{{%
\expandafter\ifx\csname l@#1\endcsname\relax
\typeout{** WARNING: IEEEtran.bst: No hyphenation pattern has been}%
\typeout{** loaded for the language `#1'. Using the pattern for}%
\typeout{** the default language instead.}%
\else
\language=\csname l@#1\endcsname
\fi
#2}}
\providecommand{\BIBdecl}{\relax}
\BIBdecl

\bibitem{kontogiannis2008research}
K.~Kontogiannis, G.~Lewis, and D.~Smith, ``{A Research Agenda for
  Service-Oriented Architecture: Research Needs for Maintenance and Evolution
  of Service-Oriented Systems},'' in \emph{Proceedings of the 2nd International
  Workshop on SOA-Based Systems Maintenance and Evolution (SOAM 2008), 12th
  European Conference on Software Maintenance and Reengineering (CSMR 2008),
  Athens, Greece}, 2008.

\bibitem{papazoglou2008service}
M.~Papazoglou, P.~Traverso, S.~Dustdar, F.~Leymann, and B.~Kramer,
  ``{Service-oriented computing: A research roadmap},'' \emph{International
  Journal of Cooperative Information Systems}, vol.~17, no.~2, pp. 223--255,
  2008.

\bibitem{arsanjani2006service}
A.~Arsanjani and A.~Allam, ``{Service-oriented modeling and architecture for
  realization of an SOA},'' in \emph{IEEE International Conference on Services
  Computing, 2006. SCC'06}, 2006, pp. 521--521.

\bibitem{Ricca2009}
F.~Ricca and A.~Marchetto, ``{A 'quick and dirty' meet-in-the-middle approach
  for migrating to SOA},'' \emph{Foundations of Software Engineering}, 2009.

\bibitem{sapmodel}
``{SAP - Enterprise Services Workplace},''
  \textit{http://esworkplace.sap.com/}.

\bibitem{ebxml}
``{OASIS - ebxml},'' \textit{http://www.oasis-open.org/home/index.php}.

\bibitem{jallinone}
``{JAllInOne},'' \textit{http://jallinone.sourceforge.net}.

\bibitem{openswing}
``{OpenSwing},'' \textit{http://oswing.sourceforge.net}.

\bibitem{jallinone-source-code}
\BIBentryALTinterwordspacing
``{JAllInOne 0.9.21}.'' [Online]. Available:
  \url{\textit{http://www.csa.iisc.ernet.in/\~{}raghavan/software/JAllInOne0.9%
.21.tar.gz}}
\BIBentrySTDinterwordspacing

\bibitem{rugaber-interleaving}
S.~Rugaber, K.~Stirewalt, and L.~M. Wills, ``The interleaving problem in
  program understanding,'' in \emph{Working Conf. in Reverse Engg.}, 1995, pp.
  166--175.

\bibitem{doxygen}
``{Doxygen},'' \textit{http://www.doxygen.org}.

\bibitem{grapheasy}
``{Graph-Easy},'' \textit{http://search.cpan.org/\~{}tels/Graph-Easy/}.

\bibitem{graphviz}
``{GraphViz},'' \textit{http://www.graphviz.org/}.

\bibitem{jip}
``{JIP - The Java Interactive Profiler},''
  \textit{http://jiprof.sourceforge.net/}.

\bibitem{Li2006}
S.~Li and L.~Tahvildari, ``A service-oriented componentization framework for
  java software systems,'' in \emph{WCRE '06: Proceedings of the 13th Working
  Conference on Reverse Engineering}.\hskip 1em plus 0.5em minus 0.4em\relax
  Washington, DC, USA: IEEE Computer Society, 2006, pp. 115--124.

\bibitem{HarmanMark2002}
M.~Harman, N.~Gold, R.~M. Hierons, and D.~Binkley, ``{Code Extraction
  Algorithms which Unify Slicing and Concept Assignment.}'' in \emph{WCRE},
  2002, pp. 11--21.

\bibitem{RajlichVaclav2002}
V.~Rajlich and N.~Wilde, ``{The Role of Concepts in Program Comprehension.}''
  in \emph{IWPC}, 2002, pp. 271--280.

\bibitem{haiduc2008use}
S.~Haiduc and A.~Marcus, ``{On the use of domain terms in source code},'' in
  \emph{Proceedings of the 2008 The 16th IEEE International Conference on
  Program Comprehension-Volume 00}, IEEE Computer Society.\hskip 1em plus 0.5em
  minus 0.4em\relax IEEE, 2008, pp. 113--122.

\bibitem{zhang2005service}
Z.~Zhang, R.~Liu, and H.~Yang, ``{Service identification and packaging in
  service oriented reengineering},'' in \emph{Proceedings of the 17th
  international conference on software engineering and knowledge engineering,
  Taipei, Taiwan, July}.\hskip 1em plus 0.5em minus 0.4em\relax Citeseer, 2005,
  pp. 14--16.

\bibitem{Zhang04}
Z.~Zhang and H.~Yang, ``Incubating services in legacy systems for architectural
  migration,'' in \emph{APSEC '04: Proceedings of the 11th Asia-Pacific
  Software Engineering Conference}.\hskip 1em plus 0.5em minus 0.4em\relax
  Washington, DC, USA: IEEE Computer Society, 2004, pp. 196--203.

\bibitem{victor91}
G.~Caldiera and V.~R. Basili, ``Identifying and qualifying reusable software
  components,'' \emph{IEEE Computer}, vol.~24, no.~2, pp. 61--70, 1991.

\bibitem{Antoniol2001}
G.~Antoniol, G.~Casazza, M.~{Di Penta}, and E.~Merlo, ``{A method to
  re-organize legacy systems via concept analysis},'' in \emph{Proceedings 9th
  International Workshop on Program Comprehension. IWPC 2001}.\hskip 1em plus
  0.5em minus 0.4em\relax IEEE Comput. Soc, 2001, pp. 281--290.

\bibitem{Kannan2008}
K.~Kannan and B.~Srivastava, ``Promoting reuse via extraction of domain
  concepts and service abstractions from design diagrams,'' in \emph{SCC '08:
  Proceedings of the 2008 IEEE International Conference on Services
  Computing}.\hskip 1em plus 0.5em minus 0.4em\relax Washington, DC, USA: IEEE
  Computer Society, 2008, pp. 265--272.

\bibitem{DelGrosso2007}
C.~{Del Grosso}, M.~{Di Penta}, and I.~G.-R. {de Guzman}, ``An approach for
  mining services in database oriented applications,'' in \emph{CSMR '07:
  Proceedings of the 11th European Conference on Software Maintenance and
  Reengineering}.\hskip 1em plus 0.5em minus 0.4em\relax Washington, DC, USA:
  IEEE Computer Society, 2007, pp. 287--296.

\bibitem{Carey2007}
M.~M. Carey and G.~C. Gannod, ``Recovering concepts from source code with
  automated concept identification,'' in \emph{ICPC '07: Proceedings of the
  15th IEEE International Conference on Program Comprehension}.\hskip 1em plus
  0.5em minus 0.4em\relax Washington, DC, USA: IEEE Computer Society, 2007, pp.
  27--36.

\bibitem{wigg97}
T.~Wiggerts, ``Using clustering algorithms in legacy systems
  remodularization,'' \emph{Working Conference on Reverse Engineering}, 1997.

\bibitem{bigg93}
T.~Biggerstaff, B.~Mitbander, and D.~Webster, ``The concept assignment problem
  in program understanding,'' \emph{ICSE}, 1993.

\bibitem{lanu93}
F.~Lanubile and G.~Visaggio, ``Function recovery based on program slicing,''
  \emph{Conference on Software Maintenance}, 1993.

\bibitem{sindhgatta2008locating}
R.~Sindhgatta and K.~Ponnalagu, ``{Locating Components Realizing Services in
  Existing Systems},'' in \emph{IEEE International Conference on Services
  Computing, 2008. SCC'08}, vol.~1, 2008.

\bibitem{wilde95}
N.~Wilde and M.~C. Scully, ``Software reconnaissance: mapping program features
  to code,'' \emph{Journal of Software Maintenance}, vol.~7, no.~1, pp. 49--62,
  1995.

\bibitem{Koshke03}
T.~Eisenbarth, R.~Koschke, and D.~Simon, ``Locating features in source code,''
  \emph{IEEE Trans. Softw. Eng.}, vol.~29, no.~3, pp. 210--224, 2003.

\bibitem{wille82}
R.~Wille, ``Restructuring lattice theory: An approach based on hierarchies of
  concepts,'' in \emph{Ordered Sets}.\hskip 1em plus 0.5em minus 0.4em\relax
  Reidel, Dordrecht-Boston, 1982, pp. 445--470.

\bibitem{Marcus2004}
A.~Marcus, A.~Sergeyev, V.~Rajlich, and J.~I. Maletic, ``{An Information
  Retrieval Approach to Concept Location in Source Code},'' \emph{WCRE}, 2004.

\bibitem{Gay2009}
G.~Gay, S.~Haiduc, A.~Marcus, and T.~Menzies, ``{On the use of relevance
  feedback in IR-based concept location},'' in \emph{2009 IEEE International
  Conference on Software Maintenance}.\hskip 1em plus 0.5em minus 0.4em\relax
  IEEE, 2009, pp. 351--360.

\bibitem{Robillard2007b}
M.~P. Robillard and G.~C. Murphy, ``{Representing concerns in source code},''
  \emph{ACM Transactions on Software Engineering and Methodology (TOSEM)},
  vol.~16, no.~1, 2007.

\end{thebibliography}

\end{document}